\documentclass[aps,prc,twocolumn,showpacs,floatfix,nofootinbib,superscriptaddress,amsmath,amssymb]{revtex4-1}

\usepackage{mathrsfs}
\usepackage{graphicx}
\usepackage{epsfig}
\usepackage{bm}
\usepackage{color}
\usepackage{float}
\usepackage{dcolumn}
\usepackage{multirow} 
\usepackage{hyperref}
\usepackage{color}
\usepackage[titletoc]{appendix}

\graphicspath{{.}{Figures/}}

\newcommand{\be}{\begin{equation}}
\newcommand{\ee}{\end{equation}}
\newcommand{\bea}{\begin{eqnarray}}
\newcommand{\eea}{\end{eqnarray}}

\begin{document}

\title{Effective field theory for nuclear vibrations with quantified
  uncertainties}

\author{E. A. Coello P\'erez} 

\affiliation{Department of Physics and Astronomy, University of
  Tennessee, Knoxville, Tennessee 37996, USA}

\author{T. Papenbrock} 

\affiliation{Department of Physics and Astronomy, University of
  Tennessee, Knoxville, Tennessee 37996, USA}

\affiliation{Physics Division, Oak Ridge National Laboratory, Oak
  Ridge, Tennessee 37831, USA}

\date{\today}

\begin{abstract}
We develop an effective field theory (EFT) for nuclear vibrations. The
key ingredients -- quadrupole degrees of freedom, rotational
invariance, and a breakdown scale around the three-phonon level -- are
taken from data.  The EFT is developed for spectra and electromagnetic
moments and transitions. We employ tools from Bayesian statistics for
the quantification of theoretical uncertainties. The EFT consistently
describes spectra and electromagnetic transitions for $^{62}$Ni,
$^{98,100}$Ru, $^{106,108}$Pd, $^{110,112,114}$Cd, and
$^{118,120,122}$Te within the theoretical uncertainties. This suggests
that these nuclei can be viewed as anharmonic vibrators.
\end{abstract}

\maketitle

\section{Introduction}

The quest for quadrupole vibrations in atomic nuclei is a long,
confusing, and opressing one. Based on the ground-breaking work by 
Bohr and Mottelson~\cite{bohr1952, bohr1953, bohr1975}, low-energy
excitations of atomic nuclei are viewed as quadrupole oscillations of
the liquid-drop surface. This approach suggests that some spherical
nuclei can be viewed as harmonic quadrupole oscillators, i.e. the
five-dimensional $U(5)$ symmetric harmonic oscillator determines their
spectra and low-lying transitions. Cadmium isotopes, for instance,
have been employed as textbook cases of vibrational
motion~\cite{bohr1975, kern1995, rowe2010}.  While corresponding
harmonic spectra (including one-, two-, and possibly three-phonon
states) were early identified in several nuclei, $B(E2)$ transition
strengths exhibit considerable deviations from the predictions of the
harmonic quadrupole oscillator, see, e.g. Refs.~\cite{lehmann1996,
  corminboeuf2000-1, kadi2003, yates2005, williams2006, garrett2007,
  bandyopadhyay2007,
  garrett2008,batchelder2009,chakraborty2011,radeck2012,batchelder2012}
for recent references to a long-standing
problem~\cite{deboer1965,bohr1975}. Proposed anharmonicities were
deemed insufficient to account for the considerable differences
between data and the harmonic
model~\cite{bes1969,bohr1975}. Particularly concerning were the
considerable variance between $B(E2)$ strengths for decays from
two-phonon states (predicted to be equal), and the relatively large
diagonal quadrupole matrix elements of low-lying $2^+$ and $4^+$ states
(predicted to vanish), see, e.g. Refs.~\cite{garrett2010, yates2012,
  garrett2012}. The observed deviations from the harmonic quadrupole
oscillator are sometimes attributed to deformation of these
thought-to-be spherical nuclei.

Based on the data it is clear that harmonic quadrupole vibrations have
not (yet) been observed in atomic nuclei. It is not clear, however,
how to understand the vibrational spectra that are evident in many
nuclei. In this paper, we revisit nuclear vibrations within an
effective field theory (EFT). The key ingredients of the EFT --
quadrupole degrees of freedom, spherical symmetry, the separation of
scale between low-lying collective excitations and a breakdown scale
at about the three-phonon level -- are consistent with data for spins
and parities of low-lying states in the nuclei we wish to describe.
The low-energy scale is approximately $\omega\approx 0.6$~MeV in
nuclei of mass number 100, while the breakdown scale $\Lambda\approx
3\omega$ is due to pairing effects and other excluded physics. At
leading order (LO), the EFT yields the harmonic quadrupole
oscillator. The breakdown scale is based on the observed proliferation
of states at about the three-phonon level, which is clearly
incompatible with the expectations from the LO Hamiltonian.  In an
EFT, corrections to the LO Hamiltonian are due to the excluded physics
beyond the breakdown scale. A power counting can be used to estimate
their size, and to systematically improve the Hamiltonian -- order by
order -- as well as transition operators. This is the program we
follow in this paper.

We note that EFTs now have a decades-old history in the physics of
nuclei. Most effort has been dedicated to an EFT of the interactions
between nucleons itself, see
Refs.~\cite{vankolck1999,bedaque2002,epelbaum2009,machleidt2011} for
reviews. Paired with {\it ab initio}
calculations~\cite{navratil2009,barrett2013,hagen2013c}, such
interactions now provide us with a model-independent approach to
atomic nuclei. Halo EFT exploits the separation of scale between
weakly-bound halo nucleons and core excitations at much higher
energy~\cite{bertulani2002,higa2008,hammer2011,ryberg2014}. The EFT
for heavy deformed nuclei~\cite{papenbrock2011,coelloperez2015}
exploits the separation of scale between low-lying rotational modes
and higher-energetic vibrations that result from the quantization of
Nambu-Goldstone modes in finite
systems~\cite{papenbrock2014,papenbrock2015}.

In this paper we also spend a considerable effort on the
quantification of theoretical uncertainties. If a theoretical result
is within the experimental uncertainties, theorists usually claim
success. However, for meaningful predictions, theoretical
uncertainties are crucial. Likewise, disagreement between theoretical
results and data can only be claimed based on the absence of overlap
between theoretical and experimental uncertainties.  Thus, the claim
that traditional vibrational models do not describe the existing data
is hard to quantify in the absence of theoretical uncertainties. This
makes uncertainty quantification particularly relevant for this work.

When it comes to theoretical uncertainties, EFTs have a key advantage
over models. The power counting immediately provides the EFT
practitioner with uncertainty estimates. Very recently, progress has
also been made toward the quantification of
uncertainties~\cite{cacciari2011,bagnaschi2015,furnstahl2015-1,furnstahl2015-2}
using Bayesian statistics. In an EFT, uncertainties can be quantified
because the (testable) expectation of ``naturalness'' can be encoded
into priors. Here, one assumes that natural-sized coefficients govern
the EFT expansion for observables. In this work, we build on these
advances and also present analytical formulas for uncertainty
quantification based on log-normal priors that are so relevant for
EFTs.

This paper is organized as follows. In Sect.~\ref{ET}, we develop the
EFT for nuclear vibrations and construct the Hamiltonian and
electromagnetic operators.  In Sect.~\ref{Uncertainty} we employ
Bayesian tools for uncertainty quantification based on the the
assumption of natural sized coefficients in the EFT expansion for
observables. We compare theoretical results with data for spectra and
for electromagnetic moments and transitions in Sect.~\ref{Comparison}
and Sect.~\ref{Transitions}, respectively.  Finally, we present our
summary in Sect.~\ref{Summary}. More detailed derivations are
relegated to the Appendix~\ref{AppUQ}.

\section{Effective theory for quadrupole vibrators}
\label{ET}
In this Section, we develop the EFT for nuclear vibrations. As our
intended audience is wide, we aim at a self-contained description. In
the following subsections we will introduce the leading-order
Hamiltonian, discuss the power counting and higher-order corrections,
and develop electromagnetic couplings and observables.

\subsection{Leading-order Hamiltonian and spectrum}
The spins and parities of low-energy spectra of even-even nuclei near
shell closures suggest these can be described in terms of quadrupole
degrees of freedom. In several cases, the spectrum resembles -- at
least at low energies -- that of a quadrupole harmonic oscillator. In
nuclei with mass number about 100, the oscillator spacing is
$\omega\approx 0.6$~MeV. The fermionic nature of the nucleus manifests
itself through pair-breaking effects wich enter at about 2--3~MeV of
excitation~\cite{dean2003}. Thus the breakdown scale will be
$\Lambda\approx 3\omega$, and for definiteness we will set
$\Lambda=3\omega$ in this work.

The boson creation and annihilation operators
$d_\mu^\dagger$ and $d_\mu$ with $\mu=-2,-1,\ldots,2$, respectively, 
fulfill the usual commutation relations
\begin{equation}
\left[ d^{\dagger}_{\mu}, d_{\nu} \right] = - \delta_{\mu}^{\nu}.
\label{phcommut}
\end{equation}
We note that $d_\mu^\dagger$ are the components of the rank-two
spherical tensor $d^\dagger$. For the general construction of
spherical tensors we also introduce the spherical rank-two tensor
$\tilde{d}$ with components
\begin{equation}
\tilde{d}_{\mu} = (-1)^{\mu} d_{-\mu}.
\label{tilded}
\end{equation}

For the construction of spherical tensors we follow
Ref.~\cite{varshalovich1988} and introduce tensor products and scalar
products.  The spherical tensor $\mathcal{I}^{(I)}$ of rank $I$
\begin{eqnarray}
\mathcal{I}^{(I)}&=&\left(\mathcal{M}^{(I_1)} \otimes \mathcal{N}^{(I_2)}\right)^{(I)}
\end{eqnarray}
results from coupling the spherical tensors $\mathcal{M}^{(I_1)}$ and
$\mathcal{N}^{(I_2)}$ of ranks $I_{1}$ and $I_{2}$ respectively. Its
components 
\begin{equation}
\mathcal{I}_{M}^{(I)} = \sum_{M_{1}M_{2}} C_{I_{1}M_{1}I_{2}M_{2}}^{IM} \mathcal{M}_{M_{1}}^{(I_1)} \mathcal{N}_{M_{2}}^{(I_2)}  
\end{equation}
are given in terms of the Clebsch-Gordan coefficients
$C_{I_{1}M_{1}I_{2}M_{2}}^{IM}$ that couple spins $I_1$ and $I_2$ to
spin $I$. Similarly, the scalar product of two spherical tensors
$\mathcal{M}^{(I)}$ and $\mathcal{N}^{(I)}$ of the same rank $I$ is
\begin{eqnarray}
\mathcal{M}^{(I)} \cdot \mathcal{N}^{(I)} &=& \sum_{\mu} (-1)^{\mu} \mathcal{M}_{\mu}^{(I)} \mathcal{N}_{-\mu}^{(I)}\\
&=& \sqrt{2I+1} \left(\mathcal{M}^{(I)} \otimes \mathcal{N}^{(I)}\right)^{(0)}.
\end{eqnarray}

There are two simple operators we need to consider. The number operator
\begin{equation}
\hat{N} \equiv d^{\dagger} \cdot \tilde{d}
\label{N}
\end{equation}
is a scalar that counts the total number of phonons $N$. The angular
momentum operator is the vector
\begin{equation}
\hat{\mathbf{I}} = \sqrt{10} \left( d^{\dagger} \otimes \tilde{d} \right)^{(1)} .
\end{equation}
Both operators conserve the number of phonons. We note that the
commutator relations
\begin{eqnarray}
\left[ \hat{I}_{\mu}, d^{\dagger}_{\nu} \right] &=& \sqrt{6} C_{2\nu 1\mu}^{2\nu+\mu} d^{\dagger}_{\nu+\mu} , \\
\left[ \hat{I}_{\mu}, \tilde{d}_{\nu} \right] &=& \sqrt{6} C_{2\nu 1\mu}^{2\nu+\mu} \tilde{d}_{\nu+\mu} , 
\end{eqnarray} 
clearly identify $d^\dagger$ and $\tilde{d}$ as spherical tensors of
rank two. In contrast, $d_\mu$ are not components of a spherical
tensor.

The Hamiltonian must be a scalar under rotation. The simplest
(i.e. quadratic in the fields $d^\dagger$ and $\tilde{d}$) Hamiltonian
is
\begin{eqnarray}
\hat{H}_{\rm LO} &=&\omega \hat{N}\nonumber\\
&=& \omega \sum_{\mu} (-1)^{\mu} d^{\dagger}_{\mu} \tilde{d}_{-\mu}\nonumber\\
&=& \omega \sum_{\mu} d^{\dagger}_{\mu} d_{\mu}.
\label{HLO}
\end{eqnarray}
Here, $\omega$ is a low-energy constant (LEC) that has to be adjusted
to data.  We note that one could also consider an operator
proportional to
\begin{equation}
d^{\dagger} \cdot d^{\dagger} + \tilde{d} \cdot \tilde{d}
\end{equation}
as part of the LO Hamiltonian. However, a Bogoliubov transformation
that introduces (quasi-)boson creation and annihilation operators
\begin{equation}
D^{\dagger}_{\mu} = u_{\mu}^{*} d^{\dagger}_{\mu} + v_{\mu}^{*} \tilde{d}_{\mu} , \qquad D_{\mu} = u_{\mu} d_{\mu} + v_{\mu} \tilde{d}^{\dagger}_{\mu}
\end{equation}
with $|u_\mu|^{2}-|v_\mu|^{2}=1$ would transform such a Hamiltonian
into the diagonal form
\begin{equation}
\hat{H}_{\rm LO} = \tilde{\omega} \left( D^{\dagger} \cdot \tilde{D} \right) . 
\end{equation}
Here $\tilde{D}$ is defined in terms of $D$ similar to
Eq.~(\ref{tilded}).  This Hamiltonian conserves the number of
(quasi-)bosons and can not be distinguished from the LO
Hamiltonian~(\ref{HLO}).

Clearly, the LO Hamiltonian of the EFT for nuclear vibrations is
equivalent to the quadrupole vibrator submodel of the Bohr
Hamiltonian~\cite{bohr1952, bohr1953, bohr1975, iachello1987, rowe2010}. We note
that the five-dimensional quadrupole oscillator exhibits an $U(5)$
symmetry. Within the EFT approach, this symmetry is a (trivial)
consequence for the choice of degrees of freedom and the quadratic LO
Hamiltonian. While this symmetry might be useful in labeling basis
states, it does not reflect symmetry properties of the nuclear
interaction between nucleons.

The energies of the LO Hamiltonian
\begin{equation}
\hat{H}_{\rm LO} | \psi \rangle = E_{\rm LO}| \psi \rangle
\end{equation}
are 
\begin{equation}
\label{ELO}
E_{\rm LO}= \omega N .
\end{equation}
For the construction of the eigenstates we follow
\citeauthor{rowe2010}, and also refer the reader to
Ref.~\cite{caprio2009b}. The eigenstates of the LO Hamiltonian can be
labeled by the quantum numbers of the symmetry subgroups in the chain
\begin{equation}
\begin{array}{ccccccccc}
{\rm U(5)} & \supset & {\rm SO(5)} & \supset & {\rm U(3)} & \supset & {\rm SO(3)} & \supset & {\rm SO(2)}\\
N & & v & & \nu & & I & & M
\end{array}.
\nonumber
\end{equation}
Here $\nu$ is a radial quantum number, $I$ and $M$ are the usual
SO(3) angular momentum and its projection onto the $z$-axis, while
the seniority $v$ is the SO(5) analog of the angular momentum. From
now on, we refer  to the SO(3) angular momentum as spin.

The ground state of the system is the phonon vacuum, denoted by
$|0\rangle$. A state with $N$ excited quanta is created from the
vacuum by the application of $N$ creation operators, coupled to
appropriate spin. Given the quantum numbers $v$ and $\nu$, the
highest-weight state is defined by
\begin{equation}
|N = v + 2\nu, v, \nu, I = 2v, M = 2v \rangle
\propto  \left(d^{\dagger} \cdot d^{\dagger}\right)^{\nu} \left(d^{\dagger}_{2}\right)^{v} |0\rangle.
\end{equation}
Here, the proportionality sign expresses the absence of proper
normalization on the right-hand side.  The remaining states with
$N=v+2\nu$ phonons can be reached from the highest-weight states by
the application of suitably defined lowering operators. This
construction is similar to the construction of SO(3) irreducible
eigenstates where one starts from the state $|I,M=I\rangle$ and
obtains the remaining states of the spin-$I$ multiplet by successive
application of the spin-lowering operator. For the LO Hamiltonian, one
finds a singlet with spin $I=2$ at the one-phonon level, a triplet
with spins $I=0,2,4$ at the two-phonon level, and a quintuplet with
spins $I=0,2,3,4,6$ at the three-phonon level. It is convenient to
determine the LEC $\omega$ from the excitation energy of the
one-phonon state.

\subsection{Power counting and NLO corrections}
\label{Powercounting}
Quadrupole excitations are the low-lying collective degrees of freedom
in even-even nuclei systems near shell closures. This picture breaks
down at higher energies $\Lambda$ when the microscopic structure of
the nucleus in terms of underlying fermionic nucleons is resolved.

In an EFT, subleading corrections to the Hamiltonian arise due to the
omitted degrees of freedom. As one can write down an unlimited number
of rotational scalars in the fields $d^\dagger$ and $\tilde{d}$, we
need a power counting (in powers of the small parameter
$\omega/\Lambda$) for the systematic construction of the EFT, order by
order. As the fields $d^\dagger$ and $\tilde{d}$ do not carry any
dimension, we introduce quadrupole coordinates $\tilde{\alpha}$ and
momenta $\pi$ as
\begin{eqnarray}
\tilde{\alpha}_{\mu} &\equiv& \sqrt{\frac{1}{2}} \ell \left(d^{\dagger}_{\mu} + \tilde{d}_{\mu} \right) , \\
\pi_{\mu} &=& i \sqrt{\frac{1}{2}} \ell^{-1} \left(d^{\dagger}_{\mu} - \tilde{d}_{\mu} \right) .
\end{eqnarray}
Here $\ell \equiv (B\omega)^{-1/2}$ is the oscillator length, and $B$
is a mass parameter. These degrees of freedom  fulfill the canonical 
commutation relations
\begin{equation}
\left[ \pi_{\mu}, \alpha_{\nu} \right] = - i\delta_{\mu}^{\nu} ,  \qquad \tilde{\alpha}_{\mu} = (-1)^{\mu} \alpha_{-\mu}. 
\end{equation}
We note that both $\tilde{\alpha}$ and $\pi$ are spherical tensors of rank
two. In terms of them, the LO Hamiltonian can be written as
\begin{equation}
\hat{H}_{\rm LO} = \frac{1}{2B} \left( \pi \cdot \pi + B^{2} \omega^{2} \tilde{\alpha} \cdot \tilde{\alpha} \right) - \frac{5}{2} \omega.
\end{equation}
Thus, the size of coordinates and momenta at the $N$-phonon level is
\begin{equation}
\tilde{\alpha} \sim \sqrt{N} \ell \qquad \pi \sim \sqrt{N} \ell^{-1}.
\label{naivescale}
\end{equation}
At the breakdown scale, we have by definition
\begin{equation}
B\omega^{2} \tilde{\alpha}^{2} \sim \Lambda , 
\quad\mbox{and}\quad \frac{\pi^{2}}{B} \sim \Lambda .
\end{equation}
Thus,
\begin{equation}
\tilde{\alpha} \sim \sqrt{\frac{\Lambda}{\omega}} \ell , \quad\mbox{and}\quad
\pi \sim \sqrt{\frac{\Lambda}{\omega}} \ell^{-1}
\label{breakscale}
\end{equation}
at the breakdown scale. 

Let us write the subleading corrections to the Hamiltonian~(\ref{HLO})
as rotationally invariant terms of the form $g_{mn} \pi^{m}
\tilde{\alpha}^{n}$, with $m+n>2$. At the breakdown scale $\Lambda$,
the energy shift due to these corrections must be so large that
$N$-phonon states cannot be distinguished from states with $N\pm 1$
phonons. Thus,
\begin{equation}
g_{mn} \pi^{m} \tilde{\alpha}^{n} \sim \omega , 
\end{equation}
and this implies
\begin{equation}
g_{mn} \sim \ell^{m-n} \left(\frac{\omega}{\Lambda}\right)^{\frac{m+n}{2}} \omega
\label{Cssize}
\end{equation}
for the natural size of these coefficients. 
When the term $g_{mn} \pi^{m} \tilde{\alpha}^{n}$ is evaluated for 
coordinates and momenta of size~(\ref{naivescale}), it scales as
\begin{equation}
g_{mn} \pi^{m} \tilde{\alpha}^{n} \sim \varepsilon^{\frac{m+n}{2}} \omega.
\label{NLOsize}
\end{equation}
Here
\begin{equation}
\label{vareps}
\varepsilon\equiv(N\omega/\Lambda)
\end{equation}
is the relevant dimensionless expansion parameter of our EFT in the
$N$-phonon level. We note that leading-order energies scale as
$\varepsilon^0\omega$.

This simple analysis suggests that terms cubic in the quadrupole
fields are the dominant subleading corrections. However, such terms
change boson number and thus enter only in second order perturbation
theory, yielding a contribution of size $\varepsilon^3\omega$. Thus,
the next-to-leading order contributions come from those terms quartic
in the quadrupole fields that preserve the boson number. They
contribute corrections of the size $\varepsilon^2\omega$.  We note
that some collective models differ from the EFT's power counting by
employing cubic terms as dominant subleading corrections,
see. e.g. Refs.~\cite{bes1969, gneuss1969, gneuss1971,
  hess1980, hess1981}. We also note that the proliferation of
higher-order terms was addressed in some models by only considering
certain combinations of operators that are symmetric under exchange of
the operators.

We can now also consider the power counting directly for the operators
$d^\dagger$ and $\tilde{d}$. When acting on states at the breakdown
scale $d^\dagger\sim\tilde{d}\sim \sqrt{\Lambda/\omega}$. Demanding
that a Hamiltonian term of the structure $\omega f_m d^m$ containing
$m$ boson operators is of the size $\omega$ at the breakdown scale
thus yields $f_m\sim (\omega/\Lambda)^{m/2}$, and the whole term
$\omega f_m d^m$ scales as $\sim\omega (\omega/\Lambda)^{m/2}$ at low
energies.

Before we continue, it is interesting to discuss an alternative -- and
less conservative -- understanding of the breakdown scale. One could
also assume that the energy corrections of the terms $g_{mn} \pi^{m}
\tilde{\alpha}^{n}$ (for $m+n>2$) are of size $\Lambda$ (and not
$\omega$) at the breakdown scale. Then, the contributions from such
terms scale as $g_{mn} \pi^{m} \tilde{\alpha}^{n} \sim
(\omega/\Lambda)^{(m+n)/2} \Lambda$ at low energies. This implies that
the off-diagonal terms with $m+n=3$ contribute an energy $\sim
\omega^2/\Lambda$ in second-order perturbation theory, and this is
equal to the contribution of the $m+n=4$ terms in first-order
perturbation theory. Such an approach would again differ from the
early approach~\cite{bes1969} because terms with four boson operators
are as important as terms with three boson operators. Compared to the
more conservative approach we are taking, this would add two
additional terms (namely $(\pi\times\pi)^{(2)}\cdot\tilde{\alpha}$ and
$(\tilde{\alpha}\times\tilde{\alpha})^{(2)}\cdot\tilde{\alpha}$) at
NLO, increasing the number of unknown LECs considerably. Such an
approach would also probably increase the breakdown scale beyond the
three-phonon level, making it difficult to identify states at higher
energies. Therefore, we did choose a more conservative -- and
physically better motivated -- power counting.

To identify the linearly independent NLO terms that conserve phonon
number, we turned to Chapter 3 of Ref.~\cite{varshalovich1988} and
determined the following three terms
\begin{eqnarray}
\hat{N}^{2} &=& \left( d^{\dagger} \cdot \tilde{d} \right)^{2}\\
\hat{\Lambda}^{2} &=& - \left( d^{\dagger} \cdot d^{\dagger} \right) \left( \tilde{d} \cdot \tilde{d} \right) + \hat{N}^{2} - 3 \hat{N}\\
\hat{I}^{2} &=& 10 \left( d^{\dagger} \otimes \tilde{d} \right)^{(1)} \cdot \left( d^{\dagger} \otimes \tilde{d} \right)^{(1)}
\end{eqnarray}
Here the operator $\hat{\Lambda}$ is the SO(5) analog of the spin
$\hat{I}$ (see, e.g., Ref.~\cite{rowe2010}). The action of these
operators on the LO states is
\begin{eqnarray}
\hat{N}^{2} | Nv\nu IM \rangle &=& N^{2} | Nv\nu IM \rangle\\
\hat{\Lambda}^{2} | Nv\nu IM \rangle &=& v(v+3) | Nv\nu IM \rangle\\
\hat{I}^{2} | Nv\nu IM \rangle &=& I(I+1) | Nv\nu IM \rangle.
\end{eqnarray}

Thus, at NLO the Hamiltonian takes the form $\hat{H}_{\rm NLO}=\hat{H}_{\rm LO} +
\hat{h}_{\rm NLO}$ with
\begin{equation}
\hat{h}_{\rm NLO} = g_{N}\hat{N}^{2} + g_{v}\hat{\Lambda}^{2} + g_{I}\hat{I}^{2}.
\label{NLOcorrect}
\end{equation}
Here, the LECs $g_N$, $g_v$, and $g_I$ have to be adjusted to data. 
The action of the NLO correction~(\ref{NLOcorrect}) on the eigenstates of
the LO Hamiltonian yields
\begin{equation}
\hat{h}_{\rm NLO} | Nv\nu IM \rangle = e_{\rm NLO} | Nv\nu IM \rangle
\end{equation}
with
\begin{eqnarray}
\label{eNLOcorr}
e_{\rm NLO} &=& g_{N} N^{2} + g_{v} v(v+3) + g_{I} I(I+1).
\end{eqnarray}
The total energy at NLO is thus
\begin{equation}
\label{ENLO}
E_{\rm NLO}=E_{\rm LO} + e_{\rm NLO}
\end{equation}

The four LECs $\omega$, $g_N$, $g_v$, and $g_I$ can be determined from
the energies of the one-phonon state and the two-phonon states. Higher
excited states would then be predictions. It is clear that the quest
for higher precision of the EFT, e.g. by including
next-to-next-to-leading order terms introduces further LECs and
requires even more data to determine the Hamiltonian. This loss of
predictive power is unsatisfactory, but it is also clear that an
approach solely based on symmetry arguments -- as proposed in this
work -- naturally leads to this state of affairs. We note that the
previous approaches~\cite{gneuss1969, gneuss1971, hess1980, hess1981}
avoid the proliferation of new coupling constants by only considering
certain combinations of higher-order terms. From the EFT's
perspective, however, such a selection does not constitute a
systematic approach.

The breakdown scale $\Lambda\approx 3\omega$ is not sufficiently large
to study contributions beyond NLO. To improve predictive capabilities,
we will quantify (rather than estimate) theoretical uncertainties.
This is done in Sect.~\ref{Uncertainty}.

\subsection{Electromagnetic couplings}
\label{EMcoup}

Our EFT deals with quadrupole degrees of freedom. As the gauge
potential $\mathbf{A}(\mathbf{r})$ is a vector field, the
electromagnetic coupling of the EFT is not obvious. We can view the
quadrupole degrees of freedom as components of a scalar field that
depends on the position coordinate $\mathbf{r}$. This view suggests to
employ $\mathbf{r}=r\mathbf{e}_r(\theta,\phi)$ (with $\mathbf{e}_r$
being the usual radial unit vector~\cite{varshalovich1988}) and to
expand the vector potential as~\cite{eisenberg1970-2}
\begin{eqnarray}
\label{expand}
\mathbf{A}(r,\Omega)&=&\sum_{JM}\sum_l A_{JM,l}j_l(kr)
\sum_{mn}C^{JM}_{lm1n}Y_{lm}(\Omega)\mathbf{e}_n \ . \nonumber
\end{eqnarray}
Here, we employed spherical basis vectors $\mathbf{e}_n$ with $n=0,\pm
1$, and $j_l$ denotes the spherical Bessel function. The spherical
wave has a momentum $k$. We note that $A_{JM,l}$ are components of a
tensor of rank $J$ for fixed $l$.

The quadrupole degrees of freedom of the EFT must couple to the
components $A_{2M,l}$, and only $l=1,2,3$ contribute due to triangular
relations on spins.  In the long wavelength approximation $kr\ll 1$,
and $j_l(kr)\propto(kr)^l$. Thus, $A_{2M,1}$ is the dominant
contribution, and we gauge
\begin{equation}
\pi_\mu\to \pi_\mu -q A_{2\mu,1} \ .
\end{equation}
Here, the charge $q$ is a LEC that needs to be adjusted to data. We
are interested in single-photon transitions and only consider terms
linearly in $\mathbf{A}$. The effective electric quadrupole operator,
resulting from gauging the LO Hamiltonian~(\ref{HLO}), is thus
\begin{eqnarray}
\label{E2LO}
\hat{Q}_{\rm LO}&=&-{q\over B}\sum_\mu (-1)^\mu A_{2\,-\mu,1}\pi_{\mu} \nonumber\\
&=& -{iq\over \sqrt{2}B\ell}
\sum_\mu (-1)^\mu A_{2\,-\mu,1}\left(d^\dagger_{\mu}-\tilde{d}_{\mu}\right) . 
\end{eqnarray}

Let us also consider higher-order corrections. Hamiltonian terms
involving two momentum operators $\pi_\mu$ and one coordinate operator
$\tilde{\alpha}_\mu$ contribute to the energy at next-to-next-to
leading order, and were beyond the NLO corrections discussed in this
section. When considering single-photon transitions, gauging
essentially replaces one of the two momentum operators by the gauge
field and couples the latter to an operator of the structure
\be
\label{gaugeNLO}
\left(\pi\times\tilde{\alpha}\right)^{(2)} \propto 
-i \left(d^\dagger\times d^\dagger - \tilde{d}\times\tilde{d}\right)^{(2)} \ . 
\ee
The EFT expectation is that this operator yields a correction of
relative size $\varepsilon^{1/2}$ to the LO operator~(\ref{E2LO}). It
induces transitions between states that differ by two phonon numbers,
and we will come back to this point after discussing non-minimal
couplings.

Let us also consider non-minimal couplings and work in the Coulomb
gauge. Then, the electric field is
$\mathbf{E}=-\partial_t\mathbf{A}=-ik\mathbf{A}$. Here, we assumed an
exponential time dependence and set the speed of light to $c=1$.  We
note that $k\approx \omega$ for transitions between states that differ
by one phonon number. The electric field has an expansion similar to
Eq.~(\ref{expand}), and the expansion coefficients fulfill
$E_{JM,l}=-ik A_{JM,l}$.  The electric field couples to the quadrupole 
operator
\bea
\label{quadmomexpand}
\hat{Q}_\mu &=& {\sqrt{2}\over \ell} Q_0 \tilde{\alpha}_\mu\nonumber\\ 
&+& {2\over \ell^2}
Q_{1}(\tilde{\alpha}\times\tilde{\alpha})^{(2)}_\mu \nonumber\\ 
&+&{2\sqrt{2}\over \ell^3}\sum_{L=0,2,4}
Q_{2L}\left(\tilde{\alpha}\times\left(\tilde{\alpha}\times\tilde{\alpha}\right)^{(L)}\right)^{(2)}_\mu \nonumber\\ 
&+& \ldots \ .  
\eea
Here, factors of the oscillator length $\ell$ have been inserted such
that the LECs $Q_0$, $Q_1$, and $Q_{2L}$ have the dimension of a
quadruple moment. The factors of $\sqrt{2}$ are inserted for
convenience. The expansion of the quadrupole moment should not be a
surprise: what is not forbidden by symmetries is allowed in an EFT. We
recall that truly ``elementary'' degrees of freedom couple to
electromagnetic gauge fields solely via minimal coupling. The EFT, however,
does not deal with ``elementary'' degrees of freedom. The quadrupole
coordinates of the EFT are effective degrees of freedom at low
energies. They are composite and describe collective effects of more
microscopic ``high-energy'' degrees of freedom that are not resolved
at the low energy scale we are interested in. The non-minimal
couplings allow us to incorporate the sub-leading electromagnetic
effects of any microscopic degrees of freedom. Based on the EFT power
counting, the natural sizes of the LECs $Q_1$ and $Q_{2L}$ are
\bea
\label{quadmomscale}
Q_1&\sim& \left({\omega\over\Lambda}\right)^{1/2} Q_0 \ , \nonumber\\
Q_{2L}&\sim& {\omega\over\Lambda} Q_0 \ .
\eea
It is useful to rewrite the expansion~(\ref{quadmomexpand}) in terms
of creation and annihilation operators. This yields
\bea
\label{quadmomexpand_d}
\hat{Q}_\mu &=& Q_0 \left(d^\dagger_\mu + \tilde{d}_\mu \right)\nonumber\\
&+& Q_{1}\left(d^\dagger \times d^\dagger + \tilde{d}\times\tilde{d} + 2d^\dagger\times\tilde{d}\right)^{(2)}_\mu \nonumber\\
&+& \sum_{L=0,2,4} Q_{2L}\bigg(
d^\dagger\times\left(d^\dagger\times d^\dagger\right)^{(L)}
+\tilde{d}\times\left(\tilde{d}\times\tilde{d}\right)^{(L)}\nonumber\\
&&+d^\dagger\times\left(d^\dagger\times \tilde{d}\right)^{(L)}
+d^\dagger\times\left(\tilde{d}\times\tilde{d}\right)^{(L)}
+\ldots\bigg)^{(2)}_\mu \nonumber\\
&+& \ldots \ .
\eea
Let us consider the right-hand-side of
Eq.~(\ref{quadmomexpand_d}). The first line is the LO term for
transitions between states that differ by one phonon number. It is
equivalent to the term~(\ref{E2LO}) obtained from gauging.  This
allows us to identify
\begin{equation}
q=\sqrt{2} B k \ell Q_0 .
\end{equation}
The second line of Eq.~(\ref{quadmomexpand_d}) is the LO term for
transitions between states that differ by two-phonon numbers, and also
determines diagonal quadrupole matrix elements. Thus, diagonal quadrupole matrix elements
are expected to be factor $\sqrt{\omega/\Lambda}$ smaller than
transition quadrupole moments between states that differ by one phonon
number. The expected finite value for diagonal quadrupole matrix elements is a
significant departure from vanishing diagonal quadrupole matrix elements
obtained for the harmonic quadrupole vibrator.  The third line has NLO
corrections (LO terms) for quadrupole transitions between states that
differ by one (three) phonon numbers. Thus, the expectations from the
harmonic quadrupole vibrator that $B(E2)$ transitions from the
two-phonon states to the one-phonon state are independent of the
initial spin are expected to suffer corrections of relative size
$\omega/\Lambda$. We note that all anharmonic corrections vanish in
the harmonic limit, i.e. for $\omega/\Lambda\to 0$.

The reduced matrix elements of a tensor operator $\hat{O}$ of rank
$\lambda$ between two states $|i\rangle$ and $|f\rangle$ are defined
as
\begin{equation}
\langle f || \hat{O} || i\rangle = \frac{\sqrt{2I_{f}+1}}{C_{I_{i}M_{i}\lambda\mu}^{I_{f}M_{f}}} \langle \beta I_{f}M_{f} | \hat{O}_{\mu} | \alpha I_{i}M_{i} \rangle .
\end{equation}
Here $\beta$ and $\alpha$ denote quantum numbers irrelevant for the
reduced matrix elements. For the transition quadrupole moments we find
the well-known LO reduced matrix elements
\bea
\label{transitionQ}
\langle 0_1^+||\hat{Q}||2_1^+\rangle_{\rm LO} &=& Q_0\langle 0_1^+||d||2_1^+\rangle =\sqrt{5}Q_0 \ , \nonumber\\
\langle 2_1^+||\hat{Q}||0_2^+\rangle_{\rm LO} &=& Q_0\langle 2_1^+||d||0_2^+\rangle = \sqrt{2}Q_0 \ , \nonumber\\
\langle 2_1^+||\hat{Q}||2_2^+\rangle_{\rm LO} &=& Q_0\langle 2_1^+||d||2_2^+\rangle = \sqrt{10}Q_0 \ , \nonumber\\
\langle 2_1^+||\hat{Q}||4_1^+\rangle_{\rm LO} &=& Q_0\langle 2_1^+||d||4_1^+\rangle = \sqrt{18}Q_0 \ ,
\eea
and uncertainty estimates are of order $Q_0\omega/\Lambda$. 

For transitions between two-phonon states we find 
\bea
\label{Qtrans2phonon}
\langle 2_2^+||\hat{Q}||0_2^+\rangle_{\rm LO} &=& 2Q_1\langle
2_2^+||\left(d^\dagger\times\tilde{d}\right)^{(2)}||0_2^+\rangle
\nonumber\\ &=& 4 Q_1\ , \nonumber\\ \langle
2_2^+||\hat{Q}||4_1^+\rangle_{\rm LO} &=& 2Q_1\langle
2_2^+||\left(d^\dagger\times\tilde{d}\right)^{(2)}||4_1^+\rangle
\nonumber\\ &=& {24\over 7} Q_1 \ , 
\eea 
and uncertainty estimates are of order $Q_1\omega/\Lambda$. 

For the diagonal quadrupole matrix elements we find the LO reduced matrix
elements
\bea
\label{staticQ}
\langle 2_1^+||\hat{Q}||2_1^+\rangle_{\rm LO} &=& 
2Q_1\langle 2_1^+||\left(d^\dagger\times\tilde{d}\right)^{(2)}||2_1^+\rangle \nonumber\\
&=& 2\sqrt{5} Q_1 \ , \nonumber\\
\langle 2_2^+||\hat{Q}||2_2^+\rangle_{\rm LO} &=& 
2Q_1\langle 2_2^+||\left(d^\dagger\times\tilde{d}\right)^{(2)}||2_2^+\rangle \nonumber\\
&=& -{{6\sqrt{5}}\over 7} Q_1 \ , \nonumber\\
\langle 4_1^+||\hat{Q}||4_1^+\rangle_{\rm LO} &=& 
2Q_1\langle 4_1^+||\left(d^\dagger\times\tilde{d}\right)^{(2)}||4_1^+\rangle \nonumber\\
&=& {6\sqrt{110}\over 7}Q_1 \ ,
\eea
and uncertainty estimates are of order $Q_1\omega/\Lambda$. Thus the
LEC $Q_1$ relates the three diagonal matrix elements~(\ref{staticQ})
and the two transition matrix elements~(\ref{Qtrans2phonon}) to each
other. This prediction of the EFT will be tested in
Sect.~\ref{Transitions}.

For the transition involving a change by two phonons we find
\bea
\label{Q2phonon}
\langle 0_1^+||\hat{Q}||2_2^+\rangle_{\rm LO} &=& Q_1\langle
0_1^+||\left(\tilde{d}\times\tilde{d}\right)^{(2)}||2_2^+\rangle
\nonumber\\ &=& \sqrt{10} Q_1\ , 
\eea
and uncertainty estimates are of order $Q_1\omega/\Lambda$. This
non-minimal correction is of the same size as the NLO
correction~(\ref{gaugeNLO}) from gauging. The combination of both
terms involves the LEC of the term from gauging and the LEC $Q_1$. As
there is only one $E2$ transition in vibrational nuclei below the
breakdown energy (i.e. the three-phonon energy), the EFT has no
predictive power for this transition beyond an estimate of its natural
size. Therefore, we will not consider it here. 

The $B(E2)$ transition strengths are given in terms of the transition
matrix elements~(\ref{transitionQ}) as
\be
\label{BE2}
B(E2,I_i\to I_f) = \frac{|\langle I_f||Q||I_i\rangle|^2}{2I_i+1}
\ee

We finally also turn to magnetic moments. In the EFT at LO, magnetic moments
are due to the vector operator
\be
\hat{\mu} = g \hat{I} \ , 
\ee
and $g$ is a LEC constant. Thus magnetic moments of states with
spin $I$ have the reduced matrix elements
\be
\label{M1}
\langle I||\hat{\mu}||I\rangle = g\sqrt{I(I+1)(2I+1)} \ .
\ee
Corrections from omitted order are of relative size
$\varepsilon^{1/2}$ [e.g. from terms such as
$(\tilde{\alpha}\times\mu)^{(1)}$]. Thus, LO magnetic moments of
$I=4$ states are a factor $\sqrt{6}$ larger than magnetic moments of
$I=2$ states. This is another testable prediction of the EFT.

It is interesting to note that anharmonic corrections to the
quadrupole operator have been considered early
on~\cite{bes1969,bohr1975}. However, \citeauthor{bes1969} related the
expansion coefficients of the quadrupole operator to those of the
Hamiltonian (also using terms cubic in the boson annihilation and
creation operators as corrections to the harmonic quadrupole
vibrator). While such an approach has fewer adjustable parameters than
the EFT we constructed, it did not yield a satisfactory description of
$^{114}$Cd. Of course, there are no symmetry arguments that would link
the expansion coefficients of non-minimal couplings and the
Hamiltonian.

Let us briefly recall the adjustable parameters. The EFT for nuclear
vibrations employs one LEC at LO [namely $\omega$ in Eq.~(\ref{ELO})]
and three additional LECs at NLO [namely $g_N$, $g_v$, and $g_I$ in
  Eq.~(\ref{eNLOcorr})] for the Hamiltonian. These LECs need to be
adjusted to the energies of four states below the three-phonon level.
Thus, LO has predictive power while NLO has predictive power only for
states at the three-phonon level. As we will see in
Sect.~\ref{Comparison}, NLO predictions for the energy of the $6_1^+$
state are more accurate than expected.

Below the three-phonon level there are four strong $E2$ transitions
($2_1^+\to 0_1^+$, $0_2^+\to 2_1^+$, $2_2^+\to 2_1^+$, $4_1^+\to
2_1^+$) that change phonon number by one unit. They require the LEC
$Q_0$ to be adjusted to data. The somewhat smaller matrix elements
that govern the two $E2$ transitions between the two-phonon states
($4_1^+\to 2_2^+$, $2_2^+\to 0_2^+$), and the three diagonal $E2$
matrix elements of the states $2_1^+$, $2_2^+$, and $4_1^+$ require
the LEC $Q_1$ to be adjusted to data. Finally, one LEC [namely $g$ in
  Eq.~(\ref{M1})] determines the three magnetic moments of the
$2_1^+$, $2_2^+$, and $4_1^+$ states. In this way, the EFT provides us
with model-independent realtiuons between observables.

\section{Quantified theoretical uncertainties}
\label{Uncertainty}

The quantification of theoretical uncertainties is of growing interest
in nuclear physics. For a wide collection of articles on this topic we
refer the reader to the \citeyear{ireland2015} focus issue and its
editorial~\cite{ireland2015}.

The power counting provides the EFT practitioner with a simple tool to
{\it estimate} theoretical uncertainties as missing contributions from
higher orders. In our case, uncertainties at LO are of the size
$\mathcal{O}(\varepsilon^2\omega)$ [as they are caused by missing NLO
contributions], while uncertainties at NLO are of the size
$\mathcal{O}(\varepsilon^3\omega)$ [due to contributions beyond
NLO]. In such estimates, one implicitly assumes that the dimensionless
coefficients in front of these order-of-magnitude estimates are of
order one.

To {\it quantify} (rather than estimate) theoretical uncertainties
requires considerable effort~\cite{carlsson2015,furnstahl2015-1}. In
this Section, we follow
Refs.~\cite{cacciari2011,bagnaschi2015,furnstahl2015-2} and employ
Bayesian statistics for uncertainty quantification. Within this
approach, theoretical uncertainties can be expressed as
degree-of-belief (DOB) intervals and have a statistical meaning. The
construction of such DOB intervals requires one to make detailed
quantitative assumptions about the behavior of omitted orders in the
power counting. As a result, theoretical predictions and uncertainties
can be confronted by data (and underlying assumptions can be verified,
or modified if required).

\subsection{Analytical results for log-normal priors}

In this Subsection we follow \citeauthor{furnstahl2015-2} and present
the formalism required for uncertainty quantification. We also
present a few analytical expressions that involve log-normal priors,
which are particularly useful when ``naturalness'' arguments are
employed in EFTs.

We are interested in uncertainty estimates for observables computed in
an EFT. The power counting, i.e.  a small ratio $\varepsilon<1$
[cf. Eq.~(\ref{vareps})] of the low-energy scale and the breakdown
scale, allows us to expand an observable $X$ as
\be
X = X_0\sum_{n=0}^\infty c_n \varepsilon^n
\ee
Here, $X_0$ sets the general scale.  In practice, the sum above can
only be computed up to and including the term involving $\varepsilon^k$. This
implies that the relative uncertainty is
\be 
\Delta_k = \sum_{n=k+1}^\infty c_n \varepsilon^n \ .  
\ee 
It is our aim to quantify the uncertainty $\Delta_k$. We are
particularly interested in quantifying the residual
\be 
\label{residual}
\Delta_k^{(M)} = \sum_{n=k+1}^{k+M} c_n \varepsilon^n \ ,  
\ee 
of the first $M$ missing terms. To quantify uncertainties, one has to
make quantitative assumptions about the distribution of the expansion
coefficients $c_n$.  A key assumption is that the expansion
coefficients are independent of each other, and assumptions about the
distribution of expansion coefficients are employed as priors.

In an EFT, the expansion coefficients are assumed to be of order
unity. The log-normal distribution
\be
\label{prior1}
{\rm pr}(c)  =  {1\over \sqrt{2\pi}\sigma c} e^{-{1\over 2}\left({\log
    c\over \sigma}\right)^2}   
\ee
is consistent with this assumption. Choosing for instance $\sigma=\log
\alpha$ (with $\alpha>1$), implies that $1/\alpha\le c\le \alpha$ with
about 68\% probability.

The expansion coefficient $c_n$ is related to the prior~(\ref{prior1})
by a second prior ${\rm pr}(c_n|c)$. We consider two examples. First,
we assume that the log-normal distributed $c$ yields a hard bound on
the size of $c_n$. Thus,
\be
\label{prior2}
{\rm pr}^{(\rm hw)}(c_n|c)  =  {1\over 2c}\Theta(c-|c_n|) \ . 
\ee 
Here $\Theta(x)$ denotes the unit step function.  The
priors~(\ref{prior1}) and (\ref{prior2}) are ``set B'' of
Ref.~\cite{furnstahl2015-2}.
Alternatively, we assume that the log-normal distributed $c$ is
related to the width of the Gaussian prior
\be
\label{prior3}
{\rm pr}^{(\rm G)}(c_n|c) = {1\over \sqrt{2\pi} c} e^{-{c_n^2\over 2c^2}} \ .
\ee
Following~\cite{furnstahl2015-2} the application of Bayes' theorem
yields a probability distribution function for the uncertainty
$\Delta$, which we write as
\be
\label{master}
p_M(\Delta|c_0,\ldots,c_k) = \frac{\int\limits_0^\infty {\rm d}c \,\, 
{\rm pr}(c) p_M(\Delta|c)
\prod\limits_{m=0}^k {\rm pr}(c_m|c)}
{\int\limits_0^\infty {\rm d}c \,\,{\rm pr}(c) \prod\limits_{m=0}^k {\rm pr}(c_m|c)} \ .
\ee
Here, the prior ${\rm pr}(c)$ is the known (or expected) pdf and ${\rm
  pr}(c_n|c)$ is the pdf for a specific expansion coefficient $c_n$
given $c$. The probability of finding an uncertainty $\Delta$ given
the priors for $c$ is
\be
\label{pdef}
p_M(\Delta|c) \equiv \left[\prod\limits_{n=k+1}^{k+M}\int\limits_{-\infty}^\infty{\rm d}c_n\,\, {\rm pr}(c_n|c)\right] \delta\left(\Delta-\Delta_k^{(M)}\right) \ .
\ee
We note that the structure of Eq.~(\ref{master}) is quite
intuitive. The numerator captures our understanding of how the
uncertainty depends on the expansion coefficients given the pdf ${\rm
  pr}(c)$, while the denominator is a normalization.

Reference~\cite{furnstahl2015-2} presents detailed discussions of
$p_M(\Delta|c_0,\ldots,c_k)$ for several combinations of priors but
does not give analytical expressions for the log-normal distributed
prior relevant for EFTs. In what follows, we derive analytical results
for the the pdf~(\ref{master}) based on the hard-wall
prior~(\ref{prior2}) for $M=1,2$. For the Gaussian
prior~(\ref{prior3}), we reduce the pdf~(\ref{master}) to single
integrations for general $M$. We hope that these formulas might be
useful also for other applications of Bayesian uncertainty
quantification in EFTs.

To make progress in computing the pdf~(\ref{pdef}), we 
rewrite the $\delta$ function as a Fourier integral
\bea
\delta\left(\Delta-\Delta_k^{(M)}\right) 
&=& {1\over 2\pi}\int\limits_{-\infty}^\infty{\rm d} t \,\,e^{i t \Delta}\prod\limits_{n=k+1}^{k+M} e^{-i t c_n \varepsilon^n} \nonumber \ .
\eea
Thus, $p_M(\Delta|c)$ is the Fourier transform of a product of Fourier
transforms
\bea
\label{pM_gen}
\lefteqn{p_M(\Delta|c) =}\nonumber\\
&{1\over 2\pi}\int\limits_{-\infty}^\infty {\rm d} t e^{i\Delta t}
\prod\limits_{n=k+1}^{k+M} \int\limits_{-\infty}^\infty {\rm d}c_n {\rm pr}(c_n|c) e^{-it\varepsilon^n c_n} \ .
\eea
We evaluate the pdf~(\ref{pM_gen}) for the Gaussian prior~(\ref{prior3}) and find
\be
\label{pMgauss}
p_M^{(\rm G)}(\Delta|c) = {1\over \sqrt{2\pi}qc} e^{-{\Delta^2\over 2q^2c^2}} \ .
\ee
Here
\be
q^2\equiv \sum_{n=k+1}^{k+M} \varepsilon^{2n} = \varepsilon^{2k+2}{1-\varepsilon^{2M}\over 1-\varepsilon^2}
\ee
depends on $M$.  Putting all together, we are left with a single
integration and can write
\bea
\label{res13}
\lefteqn{p_M^{(\rm G)}(\Delta|c_0,\ldots,c_k) = }\nonumber \\
&&{1\over \sqrt{2\pi} q}\frac
{\int\limits_0^\infty {\rm d}x\,\, x^{k+1} e^{-{1\over 2\sigma^2}(\log(x))^2}
e^{-{\gamma^2+\Delta^2/q^2 \over 2}x^2} }
{\int\limits_0^\infty {\rm d}x\,\, x^{k} e^{-{1\over 2\sigma^2}(\log(x))^2}
e^{-{\gamma^2\over 2}x^2} } \ .
\eea
In this formula, the information from the expansion coefficients
enters via
\be
\gamma^2\equiv \sum_{n=0}^k c_n^2 \ .
\ee
The numerical evaluation of the pdf~(\ref{res13}) poses no difficulty
for any value of $M$. Formula~(\ref{res13}) is one of the main results
in this Subsection.

Let us turn to the hard-wall prior~(\ref{prior2}).  For the computation of the
Fourier transform of the prior ${\rm pr}^{(\rm hw)}(c_n|c)$ we use
\be
\int\limits_{-\infty}^\infty {\rm d} c_n \,\,{\rm pr}^{(\rm hw)}(c_n|c) e^{-i t c_n \varepsilon^n}
= {\sin{(c \varepsilon^n t)}\over c \varepsilon^n t} \ , 
\ee 
and obtain the pdf for the uncertainty $\Delta$ as
\be
\label{p}
p_M^{(\rm hw)}(\Delta|c) = 
{1\over 2\pi}\int\limits_{-\infty}^\infty{\rm d} t \,\,\cos{(t \Delta)}
\prod\limits_{n=k+1}^{k+M} {\sin{(c \varepsilon^n t)}\over c\varepsilon^n t} \ .
\ee
As we will see, the integration over ${\rm d}t$ can be performed but
becomes cumbersome for $M>1$. Here, we focus on $M=1$ and present the
result for $M=2$ in the App.~\ref{AppUQ}. For $M>2$ it might be
attractive to perform the integrations numerically. In this case, two
integrations [one over ${\rm d}t$ for $p_M(\Delta|c)$ and one over
  ${\rm d}c$] remain for the computation of Eq.~(\ref{master}), and
this number is independent of $M$.

We set $M=1$ in Eq.~(\ref{p}) and obtain~\cite{gradshteyn}
\be
\label{pM1}
p_1^{(\rm hw)}(\Delta|c) = {1\over 2 c \varepsilon^{k+1}}\Theta\left(c \varepsilon^{k+1}-|\Delta|\right) \ .
\ee
This result can also be written as $p_1^{(\rm hw)}(\Delta|c) = {\rm
  pr}^{(\rm hw)}(\Delta|c)/(c \varepsilon^{k+1})$. It could also have been obtained by
direct evaluation of the ${\rm d}c_{k+1}$ integration in
Eq.~(\ref{master}) exploiting the $\delta$ function.

Let us compute $p_1^{(\rm hw)}(\Delta|c_0,\ldots,c_k)$.  We insert the pdf~(\ref{pM1})
and the priors~(\ref{prior1}) and (\ref{prior2}) into Eq.~(\ref{master}), and 
perform the integrations (see App.~\ref{AppUQ} for details). This yields
\be
\label{result1}
p_1^{(\rm hw)}(\Delta|c_0,\ldots,c_k)=
{e^{{2k+3\over 2}\sigma^2}\over 2 \varepsilon^{k+1}}
\frac
{1-\Phi\left({\sigma\over\sqrt{2}}\left(k+2 +{\log b\over \sigma^2}\right) \right)}
{1-\Phi\left({\sigma\over\sqrt{2}}\left(k+1 +{\log a\over \sigma^2}\right)\right)} \ ,  
\ee
Here, $\Phi(x)\equiv(2/\sqrt{\pi})\int_0^x {\rm d}t \exp{(-t^2)}$
denotes the error function,
\be
a\equiv \max(|c_0|,\ldots,|c_k|) ,
\ee
and 
\be
b\equiv \max\left(a,{|\Delta|\over \varepsilon^{k+1}}\right) .
\ee

Let us discuss the result~(\ref{result1}). Increasing $\Delta$ from
zero, $p_1^{(\rm hw)}(\Delta|c_0,\ldots,c_k)$ remains a constant for $b\le a$,
i.e. for $\Delta\le a \varepsilon^{k+1}$. Past this point,
$p_1^{(\rm hw)}(\Delta|c_0,\ldots,c_k)$ decays rapidly to zero as $1-\Phi$
approaches zero for increasing values of its argument.

For $k\gg 1$, we have
\be
1-\Phi(x)\approx {e^{-x^2}\over\sqrt{\pi} x} \ , 
\ee
and obtain for $b\le a$
\be
p_1^{(\rm hw)}(\Delta|c_0,\ldots,c_k) \approx {1\over 2 a \varepsilon^{k+1}} \ .
\ee
Interestingly, the same value is found if the priors~(\ref{prior1})
and (\ref{prior2}) are replaced by ``set A'' of
Ref.~\cite{furnstahl2015-2}. This sheds light on the recent
observation~\cite{furnstahl2015-2} that DOB percentages depend very
mildly on the prior as $k$ increases.

So far, we have limited our considerations to priors ${\rm pr}(c_n|c)$
that have zero mean $\overline{c_n}=0$. If one drives an EFT to
sufficiently high order, one could actually study the distribution of
the expansion coefficients $c_n$ and thereby assess the prior. As we
will see below, priors of interest to our applications have a nonzero
mean $\overline{c_n}\equiv \langle c_n\rangle\ne 0$. Thus, we need to include this 
information. 

In what follows, we assume that the priors for $c_n$ with $n\le k$
have a nonzero mean $\overline{c_n}$, but keep the priors for $c_{k+1}, c_{k+2},
\ldots$, with a zero mean (due to lack of better knowledge). Then
\bea
p_M(\Delta|c_0,\ldots,c_k)\to 
p_M\left(\Delta|c_0-\overline{c_0},\ldots,c_k-\overline{c_k}\right) \ , 
\eea
i.e. one only subtracts the mean from the coefficients $c_n$ with $n\le
k$ before inserting them into the analytical formulas.

For the Gaussian prior~(\ref{prior3}) we would also consider the
modification that the log-normal distributed $c$ is proportional (but
not equal to) the width of the Gaussian.  Thus, we introduce a scale
factor $s$ and consider the prior
\be
\label{prior3b}
{\rm pr}^{(\rm G)}(c_n|c) = {1\over \sqrt{2\pi} sc} e^{-\left({c_n\over 2sc}\right)^2} \ .
\ee
In this case, we need to replace $q\to sq$ in Eqs.~(\ref{pMgauss}) and
(\ref{res13}).

Given an interval $[a,b]$ in the domain of a pdf $p(x)$, its degree of
belief (DOB) is defined as
\begin{equation}
{\rm DOB}(a,b) = \int\limits_{a}^{b} dx\, p(x).
\end{equation}
We note that ${\rm DOB}(a,b) \leq 1$, and the DOB of an interval
represents the probability for the variable $x$ to take a value within
the interval $[a,b]$.

Our probability distributions $p_M(\Delta|c_0,\ldots,c_k)$ are
symmetric around $\Delta=0$. We define the corresponding DOB as
\begin{equation}
{\rm DOB}(-\delta, \delta) = \int\limits_{-\delta}^{\delta} {\rm d}x \, p_M\left(x| c_0,\ldots,c_k\right) . 
\end{equation}
For a fixed DOB, one can thus give the corresponding
uncertainty interval $\pm\delta$. In what follows, we will consider
${\rm DOB} = 0.68$. We note that the interval $\pm\delta$ would
correspond to the usual one-sigma uncertainty for Gaussian
distributions $p_M(\Delta|c_0,\ldots,c_k)$. Our probability
distributions~(\ref{res13}) and (\ref{result1}) are, however, not
Gaussians.

\subsection{Uncertainty quantification for energy levels}

Uncertainty quantification is a two-step procedure. First we adjust
LECs to data. Second, we quantify uncertainties based on assumptions
about the distributions of LECs.

At LO, the energy spectrum is that of a harmonic quadrupole
oscillator, see Eq.~(\ref{ELO}), and the LEC $\omega$ has to be
adjusted to data. For nuclear vibrations in the mass $A\approx 100$
region, $\omega\approx 0.6$~MeV.  Thus, the distribution of this LEC
is relatively sharp. It is neither log-normal distributed, nor is it
without a scale (i.e. log-uniform distributed). In what follows, we
fix the LEC $\omega$ for each nucleus by performing a least-square fit
of the objective function
\begin{equation}
\chi^2_{\rm LO}=\sum_s \frac{\left[E_{\rm exp}(s)-E_{\rm LO}(s)\right]^2}
{\sigma_{\rm exp}^2 +\sigma_{\rm LO}^2}. 
\end{equation}
Here, the sum is over states $s=2_1^+$, $0_2^+$, $2_2^+$, and $4_1^+$.
In the fit, the theoretical uncertainty is estimated as
\begin{equation}
\label{siglo}
\sigma_{\rm LO} = \omega \left({E_{\rm LO}(s)\over \Lambda}\right)^2 \ , 
\end{equation}
and the experimental uncertainty is neglected because $\sigma_{\rm
  exp} \ll \sigma_{\rm LO}$. 

At NLO, three new LECs ($g_N$, $g_v$, and $g_I$) enter the
determination of the energies, see Eq.~(\ref{ENLO}). Instead of
re-adjusting $\omega$ at NLO, we replace it by $\omega\to \omega
+g_\omega$, keep the value of $\omega$ at what was obtained at LO,
and adjust $g_\omega$. Thus, we rewrite
\begin{equation}
\label{ENLO2}
E_{\rm NLO} = \omega N + g_{\omega}N + g_{N} N^{2} + g_{v} v(v+3) + g_{I} I(I+1) .
\end{equation}
It is clear that the parameters $g_N$, $g_v$, $g_I$, and $g_\omega$
are expected to scale as $\omega^3/\Lambda^2$. In an EFT, one assumes
that $g_\alpha \Lambda^2/\omega^3$ (for $\alpha=N,v,I,\omega$) are of
order unity and constrained by log-normal distributions. We adjust these
coefficients to data by a minimizing the objective function
\begin{equation}
\chi^2_{\rm NLO}=\sum_s \frac{\left[E_{\rm exp}(s)-E_{\rm NLO}(s)\right]^2}
{\sigma_{\rm exp}^2 +\sigma_{\rm NLO}^2} . 
\end{equation}
Here, the employed states $s$ are as for the LO fit, but 
the theoretical uncertainty is estimated as
\begin{equation}
\label{signlo}
\sigma_{\rm NLO} = \omega \left({E_{\rm LO}(s)\over \Lambda}\right)^3 \ . 
\end{equation}
Again, the experimental uncertainty is again neglected because
$\sigma_{\rm exp} \ll \sigma_{\rm NLO}$. As we adjust four parameters
to four data points, the fit is exact.

Let us now turn to the quantification of theoretical uncertainties. We
note that simple uncertainty estimates can be based on the naive
estimates~(\ref{siglo}) and (\ref{signlo}) at LO and NLO,
respectively. For quantified uncertainties we adapt the methods of the
previous subsection to the problem at hand. 

We start with uncertainty quantification at LO. As discussed above,
the distribution for $\omega$ is a Dirac delta function, and LO
uncertainties are solely due to assumptions about the distribution of
LECs from higher orders. Thus,
\begin{equation}
p_1^{({\rm hw})}(\Delta) = \frac{e^{\frac{\sigma^{2}}{2}}}{4\varepsilon^{2}} 
\left[ 1 - \Phi \left( \frac{\sigma}{\sqrt{2}} \left[ 1 + \frac{\log(\Delta/\varepsilon^{2})}{\sigma^{2}}\right] \right) \right]
\label{hwLO}
\end{equation}
for the hard-wall prior~(\ref{prior2}), and 
\begin{equation}
p_M^{({\rm G})}(\Delta) = \frac{1}{2\pi \sigma qs} \int\limits_{0}^{\infty} dx\, e^{-\frac{\log^{2}x}{2\sigma^{2}}} e^{-\frac{\Delta^{2}x^{2}}{2q^{2}s^{2}}}
\label{GLO}
\end{equation}
for the Gaussian prior~(\ref{prior3}). Here $q^{2}\equiv
\sum_{m=k+1}^{k+M}\varepsilon^{2m}$ with $k=0$ for uncertainties due
to $M$ terms above the LO contribution. In Eq.~(\ref{hwLO}) it is
assumed that the uncertainty comes fully from the term proportional to
$\varepsilon^2$.

We now turn to uncertainty quantification at NLO. Returning to
Eq.~(\ref{ENLO2}), the NLO energy correction for the state
$|N,v,I\rangle$ is $\omega \varepsilon^2 c_2$ with
\begin{eqnarray}
c_{2} &\equiv& c_{2}(N,v,I)\nonumber\\
&=& \frac{g_{\omega} N + g_{N} N^{2} + g_{v} v(v+3) + g_{I} I(I+1)}{\varepsilon^{2} \omega},
\end{eqnarray}

Table~\ref{expcoeff} shows the resulting coefficients $c_2$ for each
state of the nuclei $^{62}$Ni, $^{98,100}$Ru, $^{106,108}$Pd,
$^{110,112,114}$Cd, and $^{118,120,122}$Te considered in this work.
These nuclei exhibit low-energy spectra that resemble a harmonic
quadrupole oscillator. All coefficients $c_2$ are of order one. Thus,
the products $\omega\varepsilon c_2$ are of natural size. Also shown
are the values of the vibrational scale $\omega$ for each nucleus and
the LEC $Q_0$ associated with the quadrupole moment, see
Sect.~\ref{transitions}. We note that these quadrupole moments are an
order of magnitude smaller than for rotational nuclei~\cite{bohr1975}.

\begin{table}[h!]
\caption{Values for the vibrational energy $\omega$ (in keV), the
  coefficients $c_{2}$ in states up to the two-phonon level, and the LEC
  $Q_{0}^{2}$ associated with the quadrupole moment (in Weisskopf
  units) for the nuclei studied in this work. } \centering
  \begin{tabular}{c|rrrrr|c}
    \hline\hline
    Nucleus & $\omega$ [keV] & $c_{2}(2_{1}^{+})$ & $c_{2}(0_{2}^{+})$ & $c_{2}(2_{2}^{+})$ & $c_{2}(4_{1}^{+})$ & $Q_{0}^{2}$ [W.U.] \\
    \hline 
    $^{62}$Ni & 1147.9 & 0.55 & -0.29 & 0.19 & 0.26 & 10.6 \\
    $^{98}$Ru & 668.1 & 1.02 & 0.57 & 0.88 & 0.83 & 27.8 \\
    $^{100}$Ru & 573.9 & 2.35 & 1.39 & 2.36 & 1.79 & 23.6 \\
    $^{106}$Pd & 541.8 & 1.80 & 1.38 & 1.36 & 1.80 & 30.4 \\
    $^{108}$Pd & 464.5 & 1.14 & 1.53 & 0.90 & 1.51 & 36.9 \\
    $^{110}$Cd & 696.7 & 1.57 & 1.32 & 1.33 & 1.56 & 21.1 \\
    $^{112}$Cd & 635.2 & 1.72 & 0.82 & 1.14 & 1.52 & 23.2 \\
    $^{114}$Cd & 578.3 & 1.72 & 0.93 & 1.23 & 1.53 & 21.8 \\
    $^{118}$Te & 582.9 & 0.83 & -0.52 & 0.19 & 0.40 & -- \\
    $^{120}$Te & 567.8 & 0.79 & 0.32 & 0.71 & 0.56 & 31.0 \\
    $^{122}$Te & 593.5 & -0.08 & 0.88 & 0.48 & 0.17 & 40.7 \\
    \hline\hline
  \end{tabular}
  \label{expcoeff}
\end{table}

\begin{figure}[h!]
\centering \includegraphics[width=0.45\textwidth]{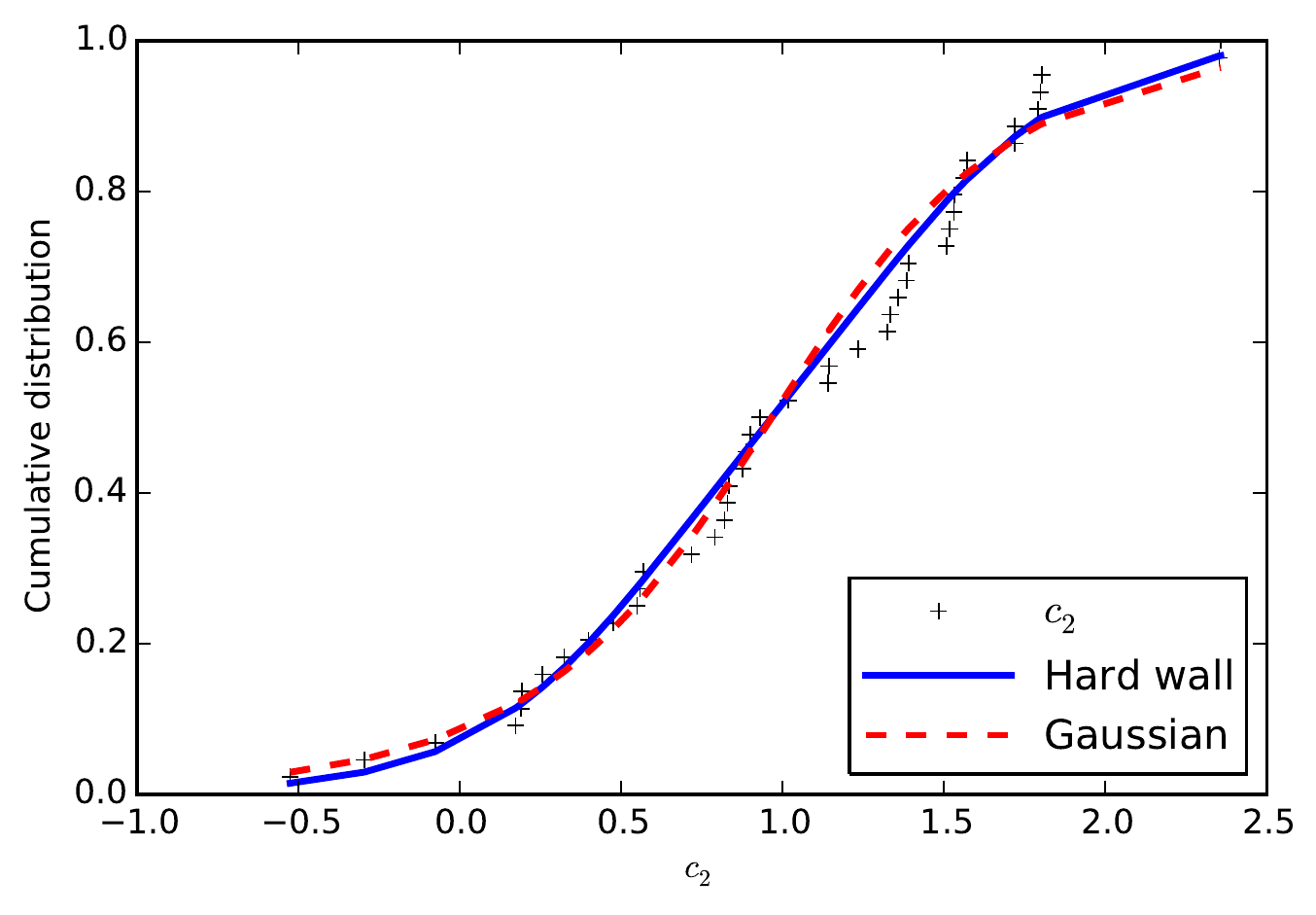}
\caption{Cumulative distribution for the $c_{2}$ coefficients for
  states up to the two-phonon level in the ensemble of all nuclei
  studied in this work. The cumulative distribution of the hard-wall
  and Gaussian priors is also shown for comparison.}
\label{cumulative}
\end{figure}

To determine a valid prior for the coefficients $c_2$ we turn to the
distribution of the coefficients $c_2$ for an ensemble consisting of
one-phonon and two-phonon states in the nuclei we study. The
cumulative distribution is shown in Fig.~\ref{cumulative}. It is well
approximated by a Gaussian prior~(\ref{prior3}) with parameter
$s\approx0.65$, or by a hard-wall prior~(\ref{prior2}), once the mean
is shifted from zero to $\mu\approx 1$.  We note that the cumulative
distribution is practically unchanged when $c_2$ values from three-phonon
states are included in the analysis. We employ $\sigma=\log{(3/2)}$ in
the log-normal prior~(\ref{prior1}).

Finally we turn to uncertainty quantification at NLO for individual nuclei.  
For the hard-wall prior we find
\begin{equation}
p_1^{({\rm hw})} \left(\Delta | c_{2}\right) 
= \frac{e^{\frac{3\sigma^{2}}{2}}}{2\varepsilon^{3}} 
\frac{1 - \Phi \left( \frac{\sigma}{\sqrt{2}} \left[ 2 + \frac{\log(\kappa)}{\sigma^{2}}\right] \right)}
{1 - \Phi \left( \frac{\sigma}{\sqrt{2}} \left[ 1 + \frac{\log(|c_{2}'|)}{\sigma^{2}}\right] \right)} .
\label{hwNLO}
\end{equation}
Here $\kappa\equiv \max(|c_{2}'|, \Delta/\varepsilon^{3})$ and $c_{2}'\equiv
c_{2}-\overline{c_2}$. For the Gaussian prior we find
\begin{equation}
p_M^{({\rm G})}\left(\Delta | c_{2}\right) = 
\frac{\int\limits_{0}^{\infty} dx\, x e^{-\frac{\log^{2}x}{2\sigma^{2}}} 
e^{-\frac{\left(c_{2}'^{2} + \Delta^{2}/q^{2}\right)x^{2}}{2s^{2}}}}{\sqrt{2\pi}qs 
\int\limits_{0}^{\infty} dx\, e^{-\frac{\log^{2}x}{2\sigma^{2}}} e^{-\frac{c_{2}'^{2}x^{2}}{2s^{2}}}}.
\label{GNLO}
\end{equation}

\begin{figure}[h!]
\centering
\includegraphics[width=0.45\textwidth]{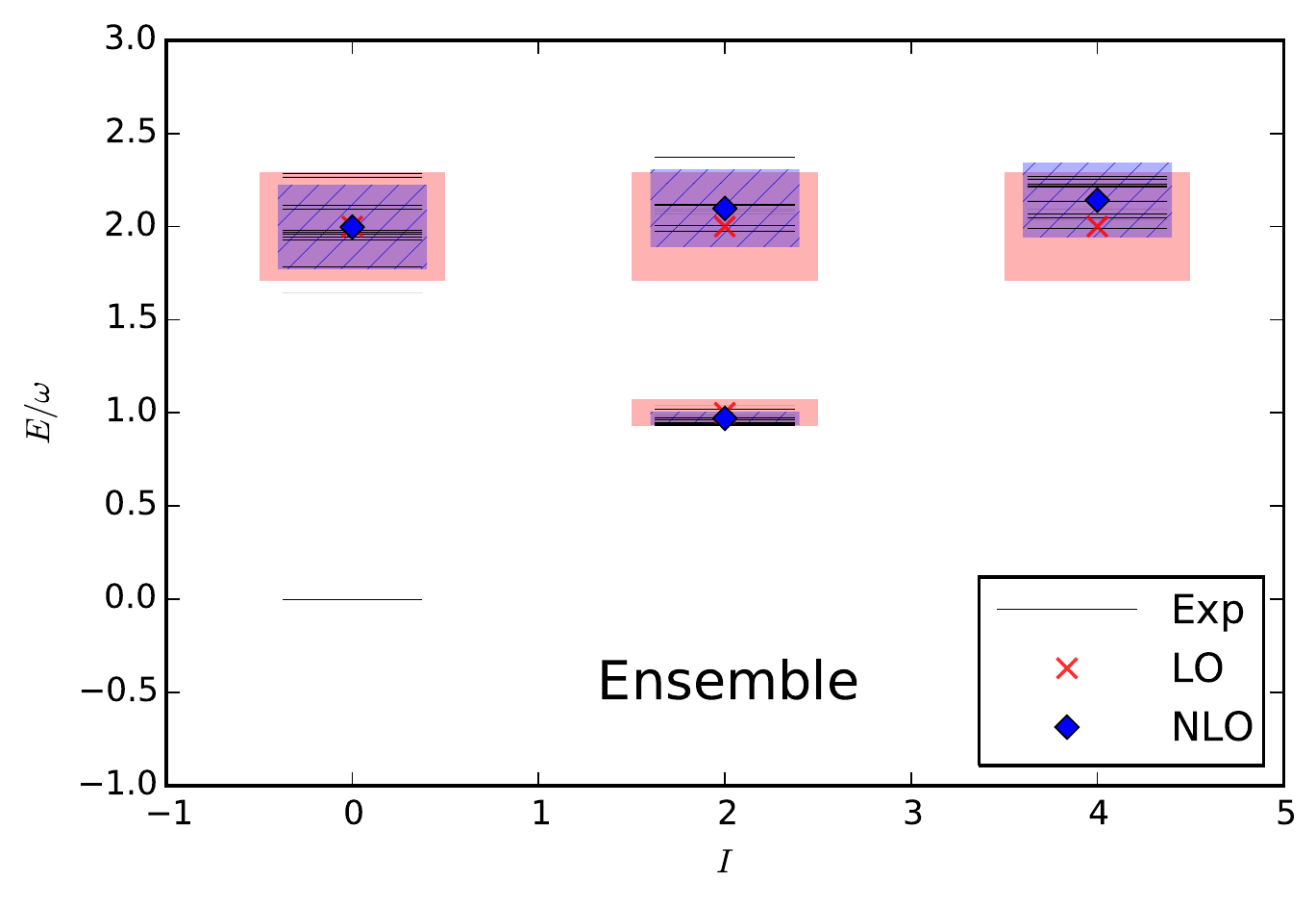}
\caption{Comparison between the normalized energies $E/\omega$ of the
  one- and two-phonon states as a function of spin $I$ in the ensemble
  of the nuclei studied in this work. Experimental energies are shown
  as thick black lines. 68\% DOB intervals are shown as shaded areas
  (plain for uncertainty quantification at LO and hatched at NLO).}
\label{ensembleE}
\end{figure}

In the determination of the prior, we employed an ensemble of
nuclei. To assess the consistency of this approach, and to verify the
statistical interpretation of the quantified uncertainties, we compare
EFT predictions for the one-phonon and two-phonon states of these
nuclei.  To do so, we first normalize the energies by dividing them by
the nucleus-dependent $\omega$, and then perform $\chi^{2}$ fits at LO
and NLO. The results are shown in Figure~\ref{ensembleE}.
Experimental data, LO calculations and NLO calculations are shown as
black lines, red crosses and blue diamonds, respectively. The
theoretical uncertainty at each order, displayed as a shaded area of
the corresponding color, are 68\% DOB intervals obtained with the
Gaussian prior. We note that 82\% of the 44 one- and two-phonon states
lie within the NLO theoretical uncertainty. This is within one sigma
($1/\sqrt{44}\approx 15$\%) of the expected 68\% for the ensemble
size. Thus, the statistical interpretation of our DOB intervals is
consistent for the energies.

\subsection{Uncertainty quantification for quadrupole moments}
We quantify uncertainties for LO transition quadrupole moments as
follows.  The expansion for these matrix elements is
\begin{equation}
\label{fiexp}
\langle f || Q || i \rangle = \langle f || Q || i \rangle_{\rm LO}\left(1 + 
\sum_{i=1} c_{i} \varepsilon^{i} \right) , 
\end{equation}
and coefficients $c_{i}$ that are expected to be of order one. The
expansion for the $B(E2)$ transition strength~(\ref{BE2}) is obtained
from the expansion~(\ref{fiexp}) of the corresponding matrix element.
We quantify uncertainties for these matrix elements and transition
strengths based on Eq.~(\ref{GLO}) with $s=1$ and compute 68\% DOB
intervals.

To summarize this Section, we have derived analytical formulas for
uncertainty quantifiaction based on log-normal priors. For uncertainty
quantification of LO results for energies and matrix elements we
employ Eq.~(\ref{GLO}) with $s=1$ and compute 68\% DOB intervals. For
uncertainty quantification at NLO for energies, we confirmed that the
prior for the employed expansion coefficients is based on data from an
ensemble of vibrational nuclei. Based on this ensemble,
Eqs.~(\ref{hwNLO}) and (\ref{GNLO}) describe the distribution of
uncertainties. These are then used for the computation of 68\% DOB
intervals.

\section{Energy spectra with quantified uncertainties}
\label{Comparison}
To test the EFT, we compare the low-energy spectra and reduced
transition probabilities of the nuclei $^{62}$Ni, $^{98,100}$Ru,
$^{106,108}$Pd, $^{110,112,114}$Cd, and $^{118,120,122}$Te against LO
and NLO results. We consider nuclei in which the ratio of energies
$E(4_{1}^{+})/E(2_{1}^{+})\approx 2$, states with the spins of the
two-phonon triplet are at about $2E(2_{1}^{+})$, and states with the
spins of the three-phonon quintuplet are around $3E(2_{1}^{+})$.
First, we discuss the description of the energy spectra by the
EFT. The LECs required for such description were obtained from
$\chi^{2}$ fits at LO and NLO, with a breakdown scale set to $\Lambda
= 3\omega$, based on the appearance of states that cannot be
identified with harmonic quadrupole excitations. 

The low-lying spectrum of $^{62}$Ni exhibits states with the spins and
energies of a harmonic quadrupole vibrator up to the three-phonon level,
making this nucleus a candidate for low-energy vibrational
behavior. The breakdown of vibrational motion at the three-phonon level
agrees with the results and discussion for this nucleus presented in
Ref.~\cite{chakraborty2011}, where shell model calculations with a
$^{40}$Ca core were required to simultaneously describe the energies
and electromagnetic properties of some multi-phonon candidates. Similar
results for this and other nickel isotopes~\cite{kenn2001, orce2008},
suggest that intruder configurations need to be taken into account in
in a microscopic description of spectra and electromagnetic properties of
the low-lying states in these nuclei. 

\begin{figure}[h!]
\centering
\includegraphics[width=0.45\textwidth]{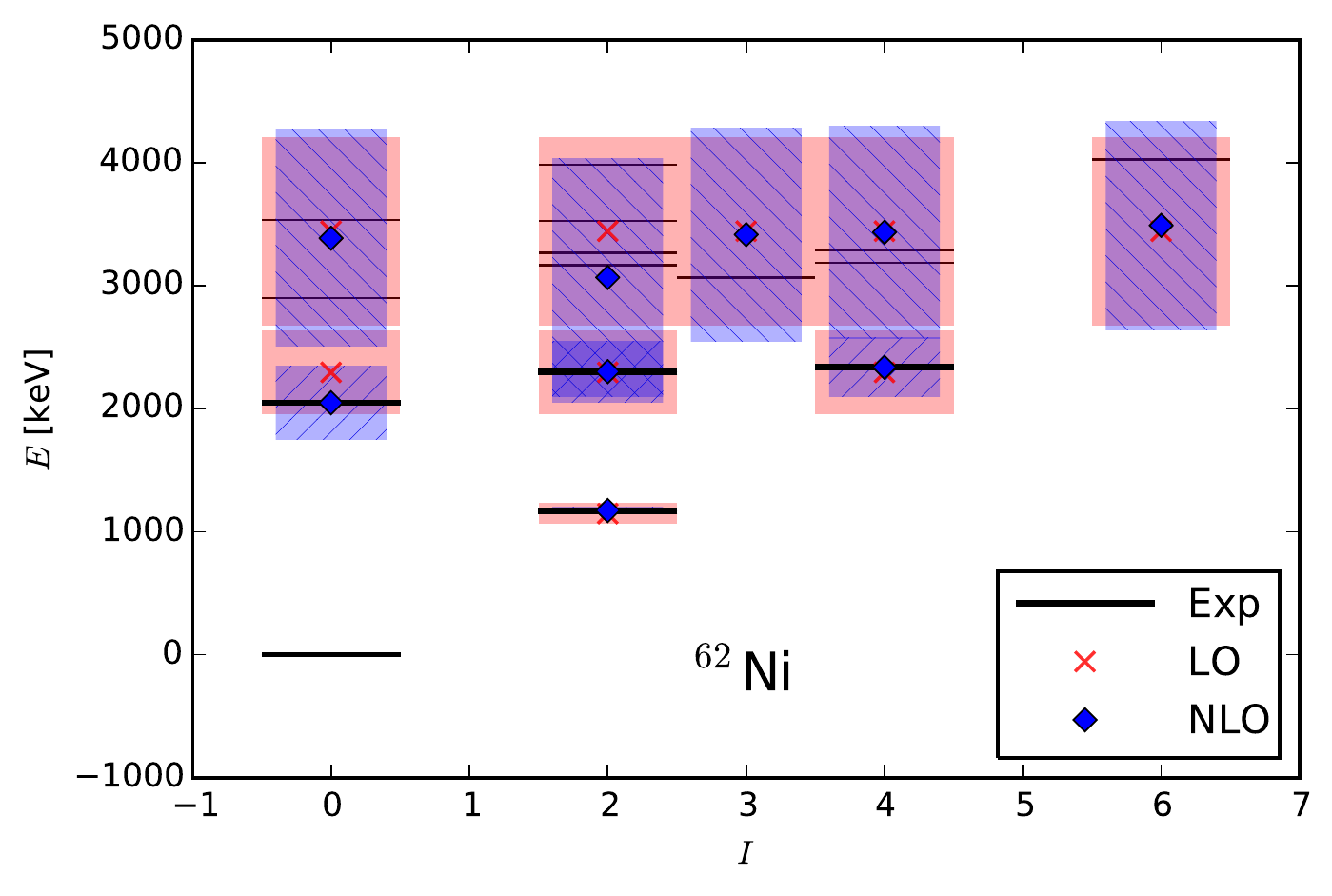}
\caption{Partial energy spectrum of $^{62}$Ni up to the three-phonon
  level. Experimental data~\cite{nichols2012} is compared to LO and
  NLO calculations of the effective theory. States up to the two-phonon
  level are shown as thick black lines. Theoretical uncertainties are
  shown as 68\% DOB intervals.}
\label{ni}
\end{figure}

Figure~\ref{ni} shows the comparison between experimental data taken
from Ref.~\cite{nichols2012}, LO and NLO calculations for energies up
to the three-phonon level for this nucleus. States up to the
two-phonon level are shown as thick black lines, while states above
them are shown as thin lines whenever a definite spin assignment have
been established (consequently, some of the nuclei studied in this
work exhibit a higher density of states above the two-phonon level
than the one displayed in the figures). The uncertainty at each order
is shown as 68\% DOB areas. The increased level density above the
two-phonon states is consistent with our identification of the
breakdown scale at about the three-phonon level. Below the breakdown
level, the description of the experimental data is improved order by
order. We note that the LO and NLO predictions for three-phonon
energies are relatively close.

Let us make three more comments that apply to $^{62}$Ni and the other
nuclei studied in what follows.  First, LO predictions are consistent
with data within the quantified theoretical uncertainties.  Second, we
note that the energies up to the two-phonon states are accurately
described at NLO, because the EFT Hamiltonian exhibits four adjustable
LECs. Thus, EFT predictions are accurate (they agree with data) yet
not very precise (theoretical uncertainties are considerable).  The
comparison of LO and NLO results shows the convergence properties of
the EFT. Third, we also note that the prediction for the $I=6$
three-phonon state is quite accurate. It thus seems that the breakdown
scale for yrast states could be higher than for the other states. This
is presumably due to the lower level density of high-spin states.

\begin{figure}[h!]
\centering
\includegraphics[width=0.45\textwidth]{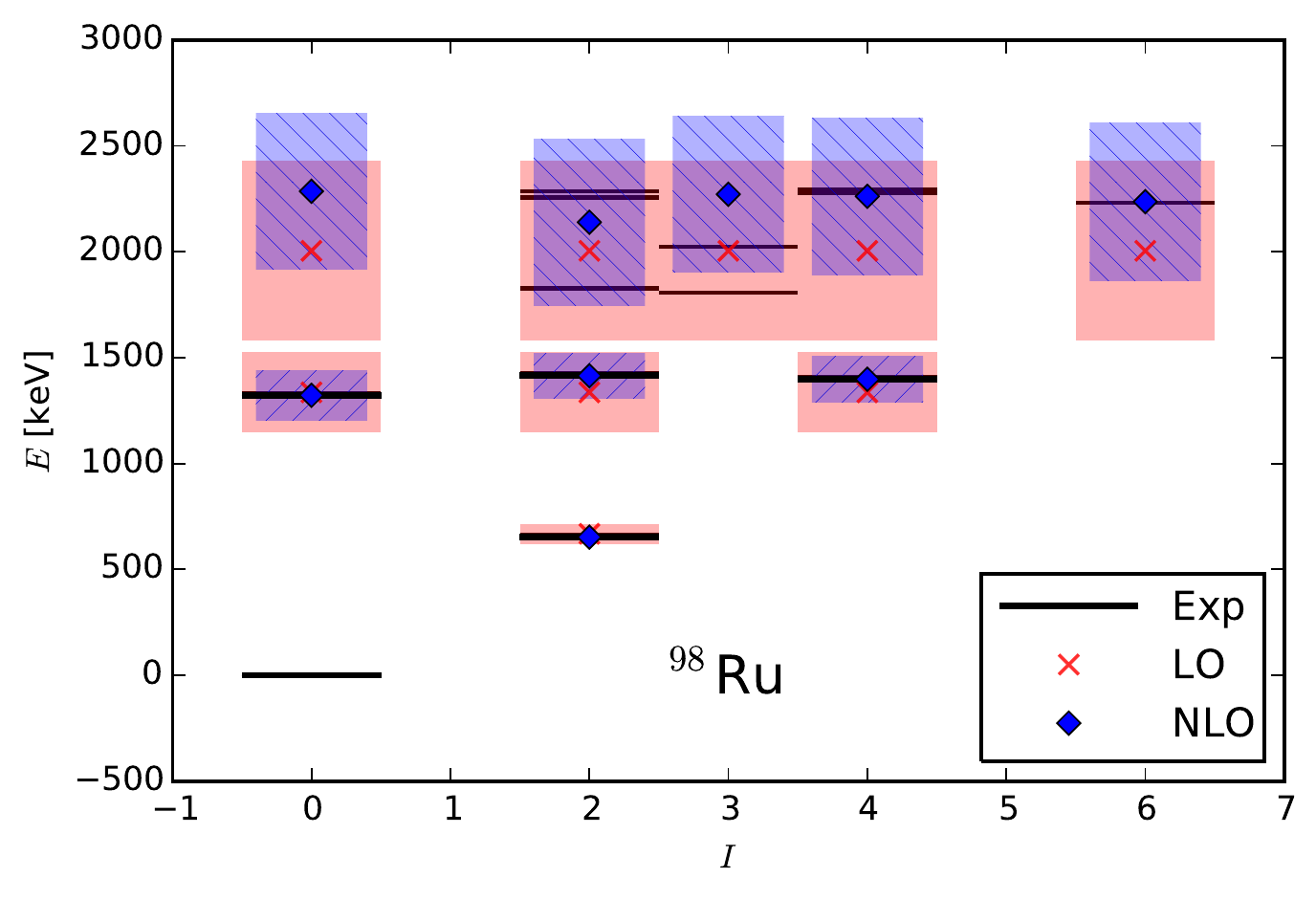}
\includegraphics[width=0.45\textwidth]{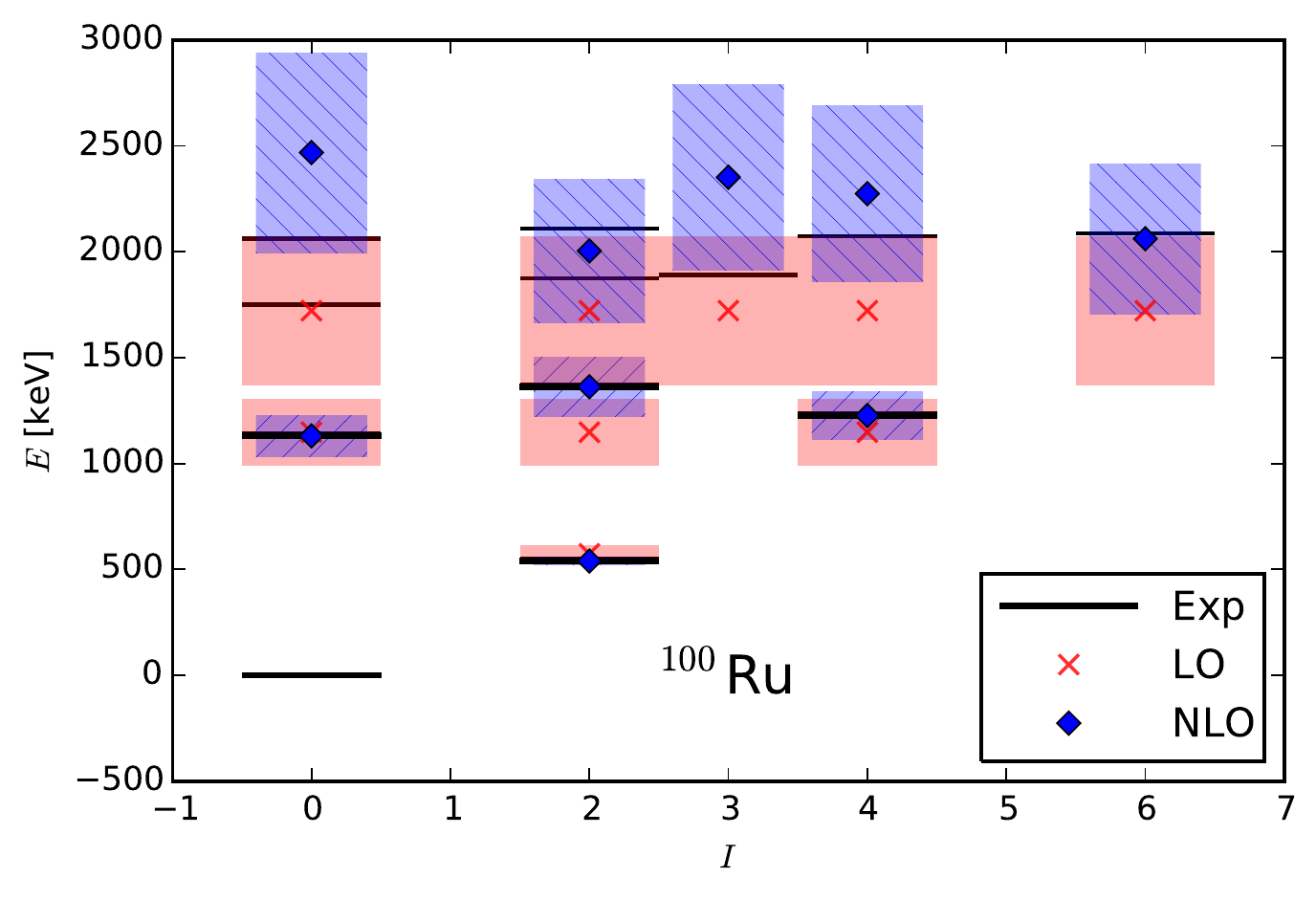}
\caption{Partial energy spectrum of $^{98}$Ru (top) and $^{100}$Ru
  (bottom) up to the three-phonon level. Experimental
  data~\cite{singh2003, singh2008} is compared to LO and NLO
  calculations of the effective theory. States up to the two-phonon
  level are shown as thick black lines. Theoretical uncertainties are
  shown as 68\% DOB intervals.}
\label{ru}
\end{figure}

Figure~\ref{ru} compares the energy spectrum of $^{98}$Ru (top) and
$^{100}$Ru (bottom) and our calculations. Again, the breakdown scale
seems properly identified.  We note that the differences between LO
and NLO predictions for three-phonon levels are considerable.

The ruthenium isotopes near the $N=50$ shell closure appear to undergo
a transition from spherical to triaxial shapes, based on the behavior
of the ratio $R_{4/2}\equiv E(4_1^+)/E(2_1^+)$ with increasing neutron
number~\cite{cakirli2004}. From this chain, $^{98}$Ru is the first
isotope expected to exhibit collective behavior based on its ratio of
energies $R_{4/2}\approx 2$. Its low-energy spectrum exhibits
vibrational-like excitations, with several non-vibrational states
above the two-phonon level. Experimental energies were taken from
Ref~\cite{singh2003}. For $^{100}$Ru, experimental data were taken
from Ref.~\cite{singh2008}. Shell model calculations with neutrons
promoted across the $N=50$ shell gap reveals the importance of single
particle motion in these isotopic chain~\cite{kharraja1999,
  timar2000}. As mentioned before, ruthenium isotopes transit from
spherical to triaxial shapes as the neutron number increase. Larger
deviations from the harmonic behavior in $^{100}$Ru suggest that
deviations from the spherical shape are larger than in $^{98}$Ru.

\begin{figure}[h!]
\centering
\includegraphics[width=0.45\textwidth]{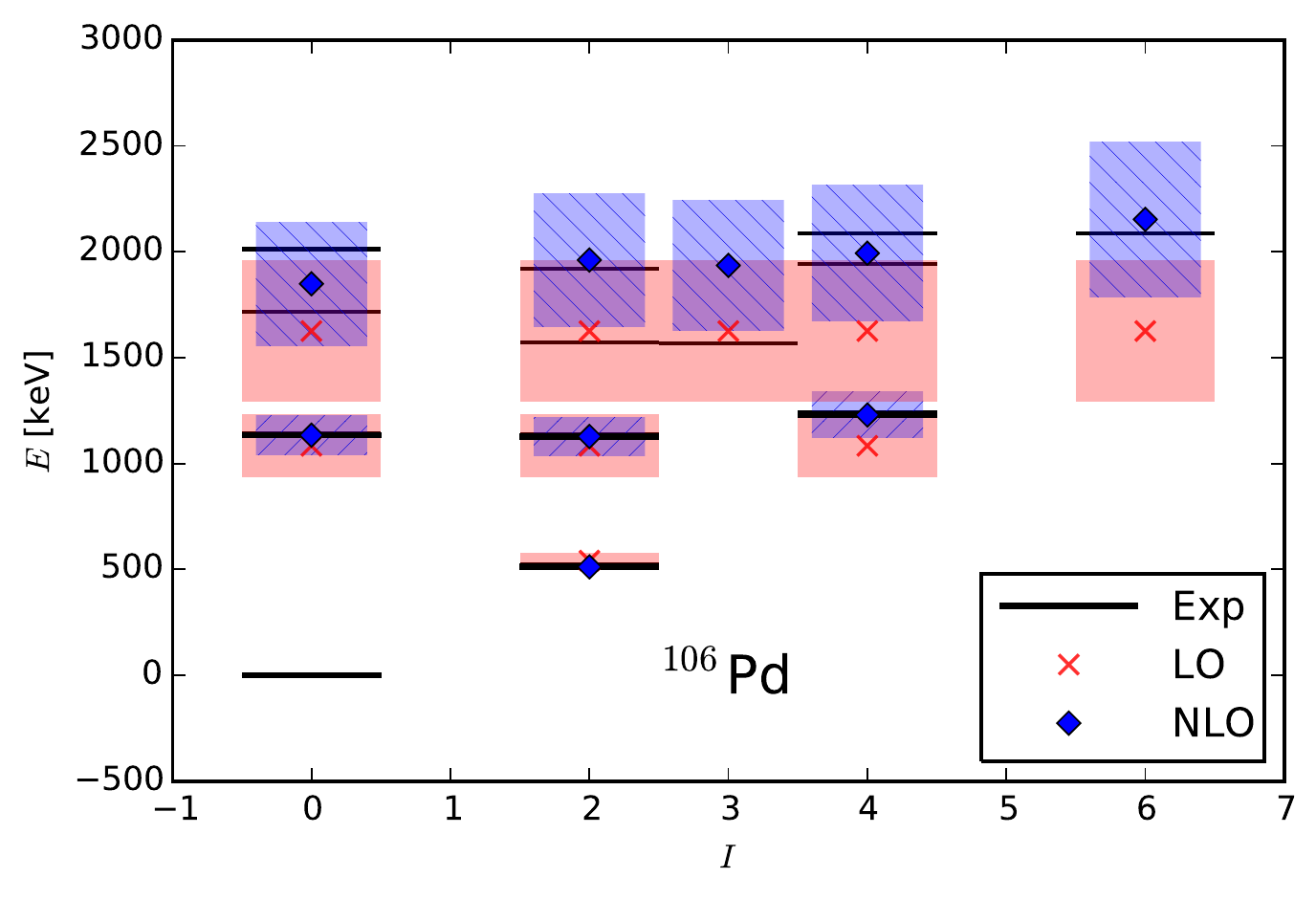}
\includegraphics[width=0.45\textwidth]{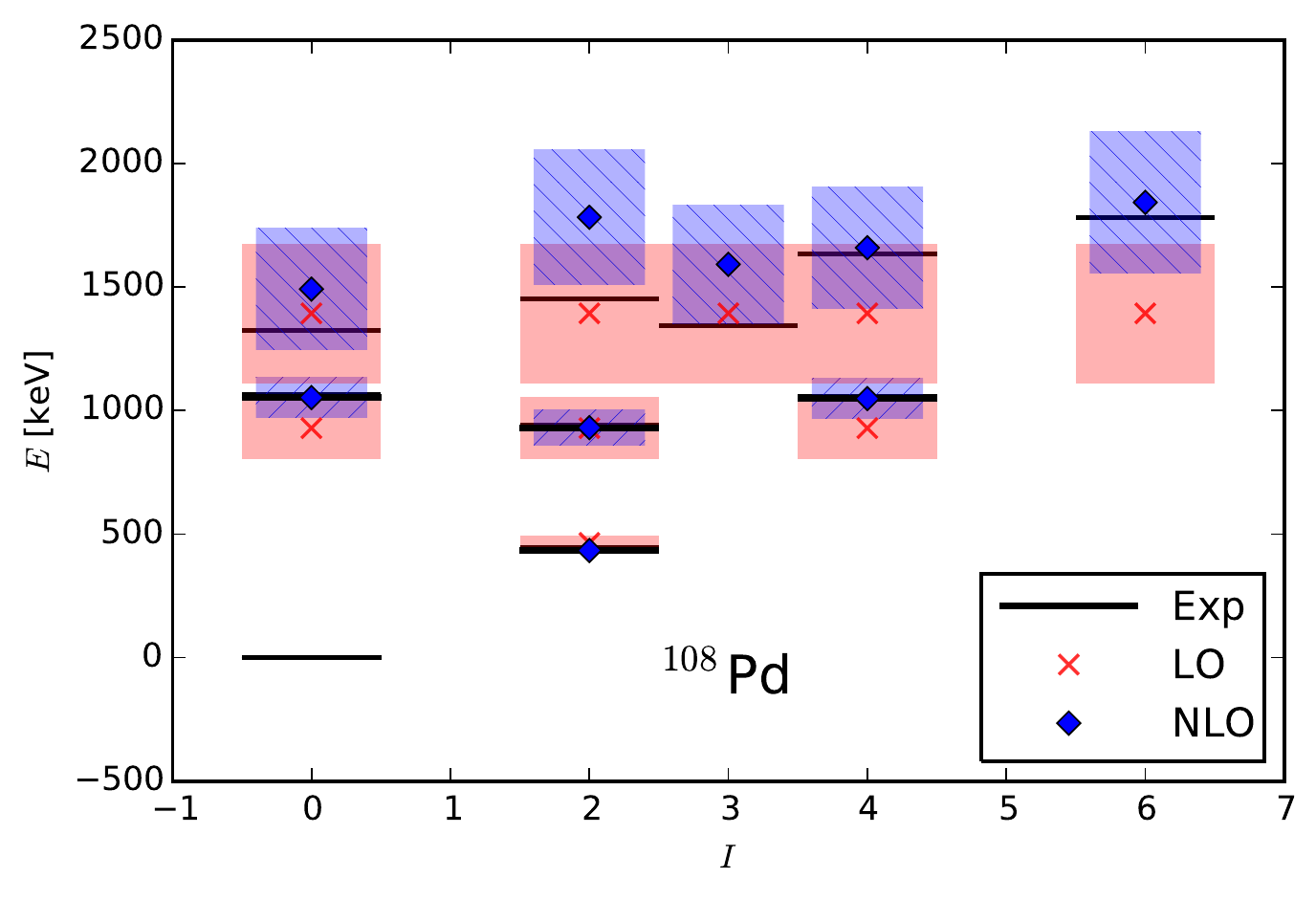}
\caption{Partial energy spectrum of $^{106}$Pd (top) and $^{108}$Pd
  (bottom) up to the three-phonon level. Experimental
  data~\cite{defrenne2008, blachot2000} is compared to LO and NLO
  calculations of the effective theory. States up to the two-phonon
  level are shown as thick black lines. Theoretical uncertainties are
  shown as 68\% DOB intervals.}
\label{pd}
\end{figure}

The energy spectra of $^{106}$Pd and $^{108}$Pd are compared against LO
and NLO calculations in the top and bottom of Figure~\ref{pd}
respectively. In $^{108}$Pd there are fewer levels around the
three-phonon states. The considerable deviations of the $I=0,2$
three-phonon energies from NLO predictions -- consistent with the
theoretical uncertainties -- nevertheless suggests that the breakdown
scale has been identified correctly.

The energy spectra and enhanced transitions probabilities for decays
from the low-lying states in palladium isotopes, assumed to be
spherical, suggest vibrational motion in these systems. For $^{106}$Pd
and $^{108}$Pd, experimental data was taken from
Ref.~\cite{defrenne2008} and Ref.~\cite{blachot2000} respectively.
Single particle states have been suggested for
$^{108}$Pd~\cite{regan1997}. The palladium isotopes exhibit ratios
$R_{4/2}\approx 2.4$ and
$B(E2;4_1^+\downarrow)/B(E2;2_1^+\downarrow)\approx 1.6$. These
quantities, in addition to the large diagonal quadrupole matrix
elements for states up to the two-phonon level in palladium
isotopes~\cite{svensson1995}, strongly suggest that the deviation from
the harmonic oscillator behavior in these systems is considerable.

\begin{figure}[h!]
\centering
\includegraphics[width=0.45\textwidth]{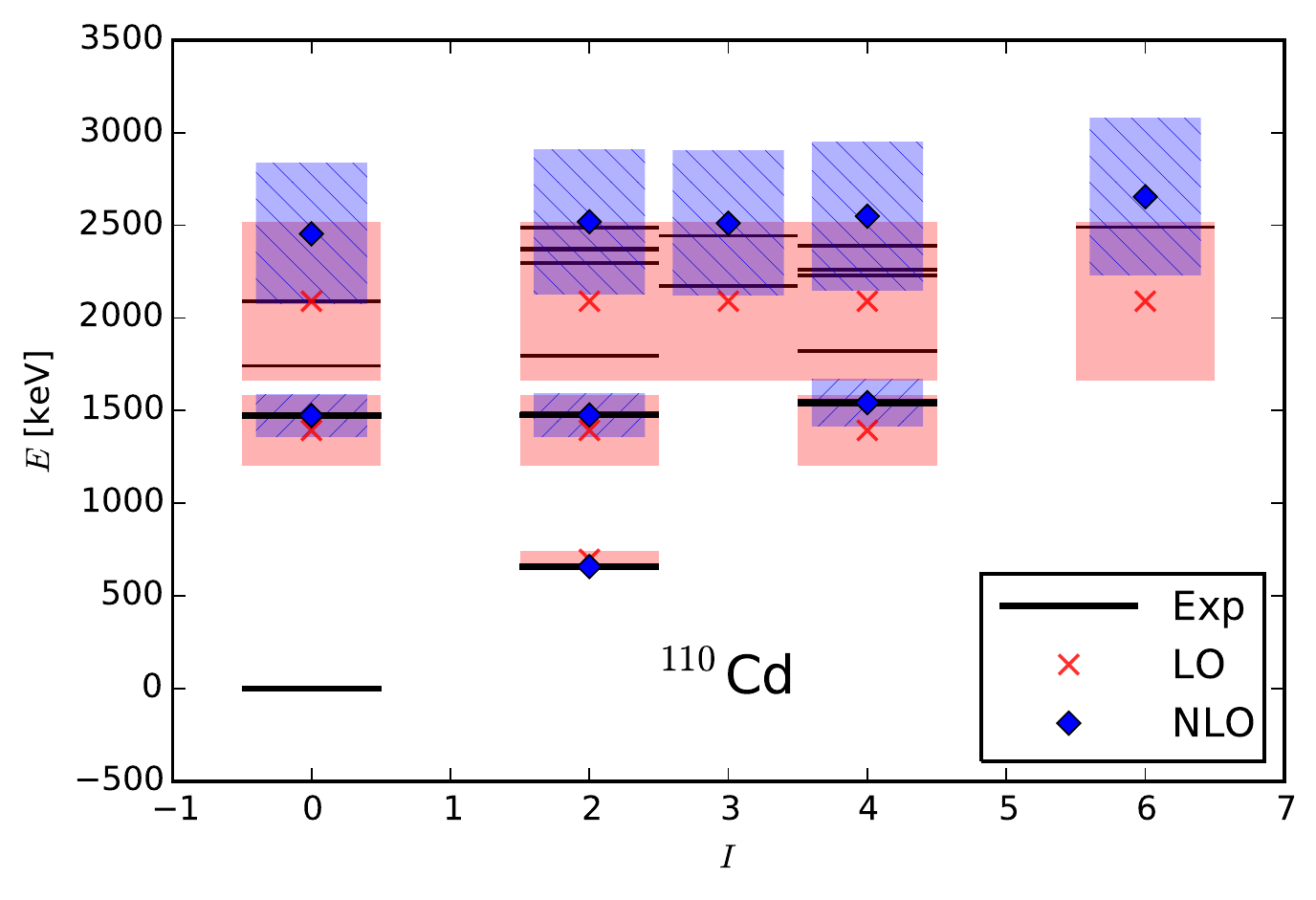}
\includegraphics[width=0.45\textwidth]{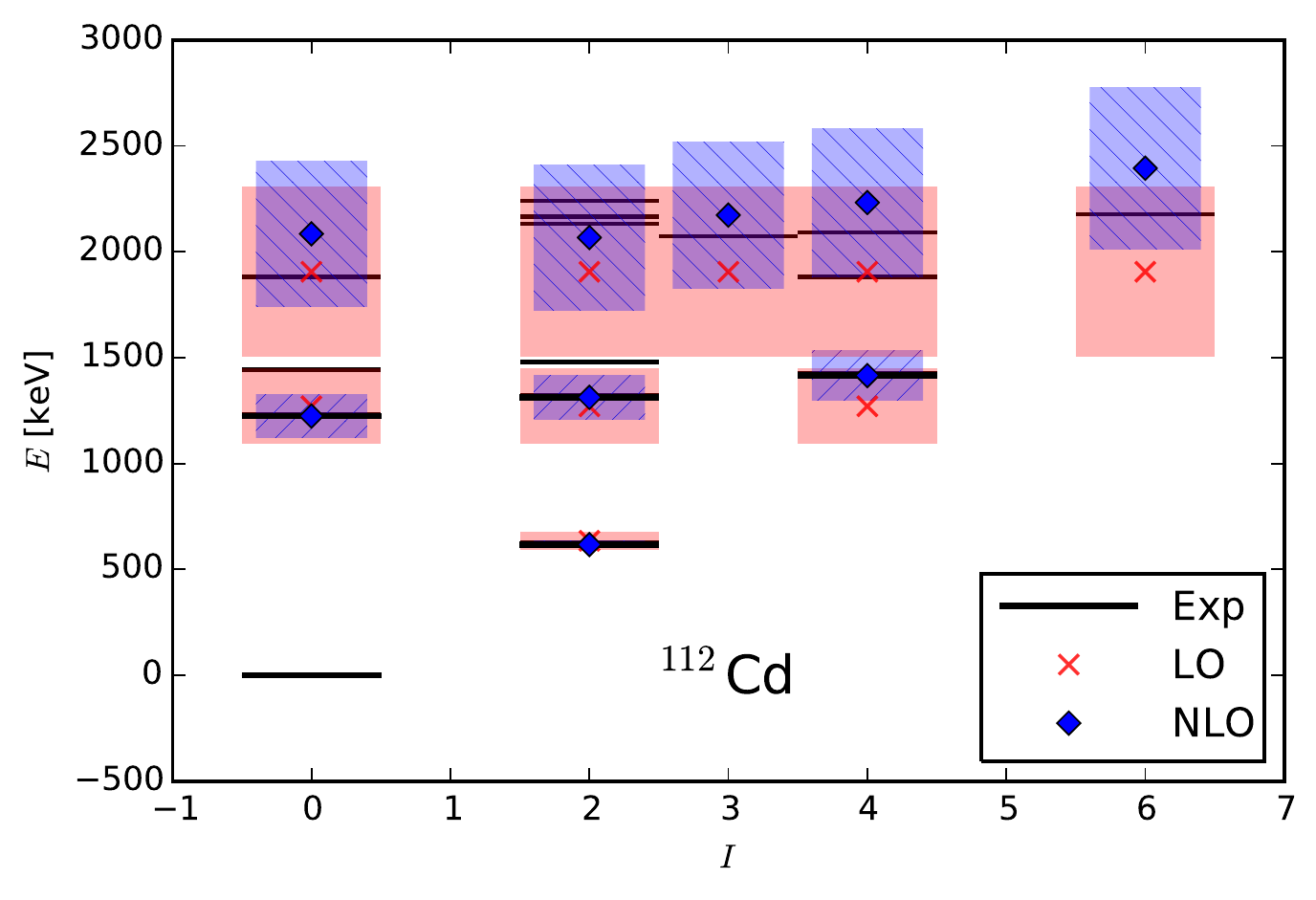}
\includegraphics[width=0.45\textwidth]{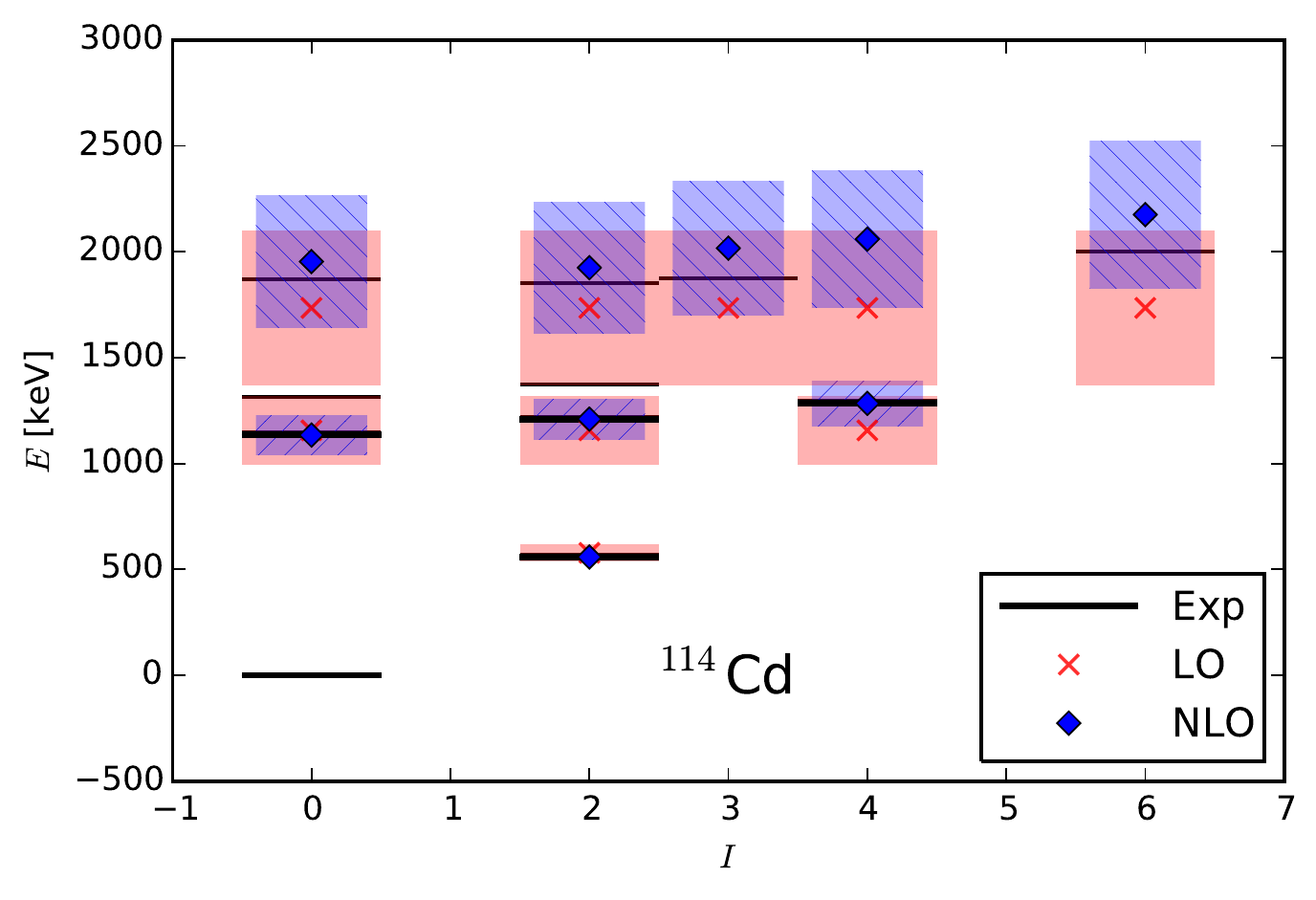}
\caption{Partial energy spectrum of $^{110}$Cd (top), $^{112}$Cd
  (middle) and $^{114}$Cd (bottom) up to the three-phonon
  level. Experimental data~\cite{gurdal2012, defrenne1996,
    blachot2012} is compared to LO and NLO calculations of the
  effective theory. States up to the two-phonon level are shown as
  thick black lines. Theoretical uncertainties are shown as 68\% DOB
  intervals.}
\label{cd}
\end{figure}

Figure~\ref{cd} compares experimental spectra of cadmium isotopes with
LO and NLO order results from EFT. We note that the deviations from
expectations for the harmonic quadrupole vibrator are pronounced in
these isotopes, with additional energy levels just above the
two-phonon states. We also note that the energies of the three-phonon
$6_1^+$ states deviate stronger from EFT predictions than for the other
nuclei we consider in this work. In these nuclei, the breakdown scale
for vibrations is clearly low. From the EFTs perspective anharmonic
corrections are expected to be most significant.

\begin{figure}[h!]
\centering
\includegraphics[width=0.45\textwidth]{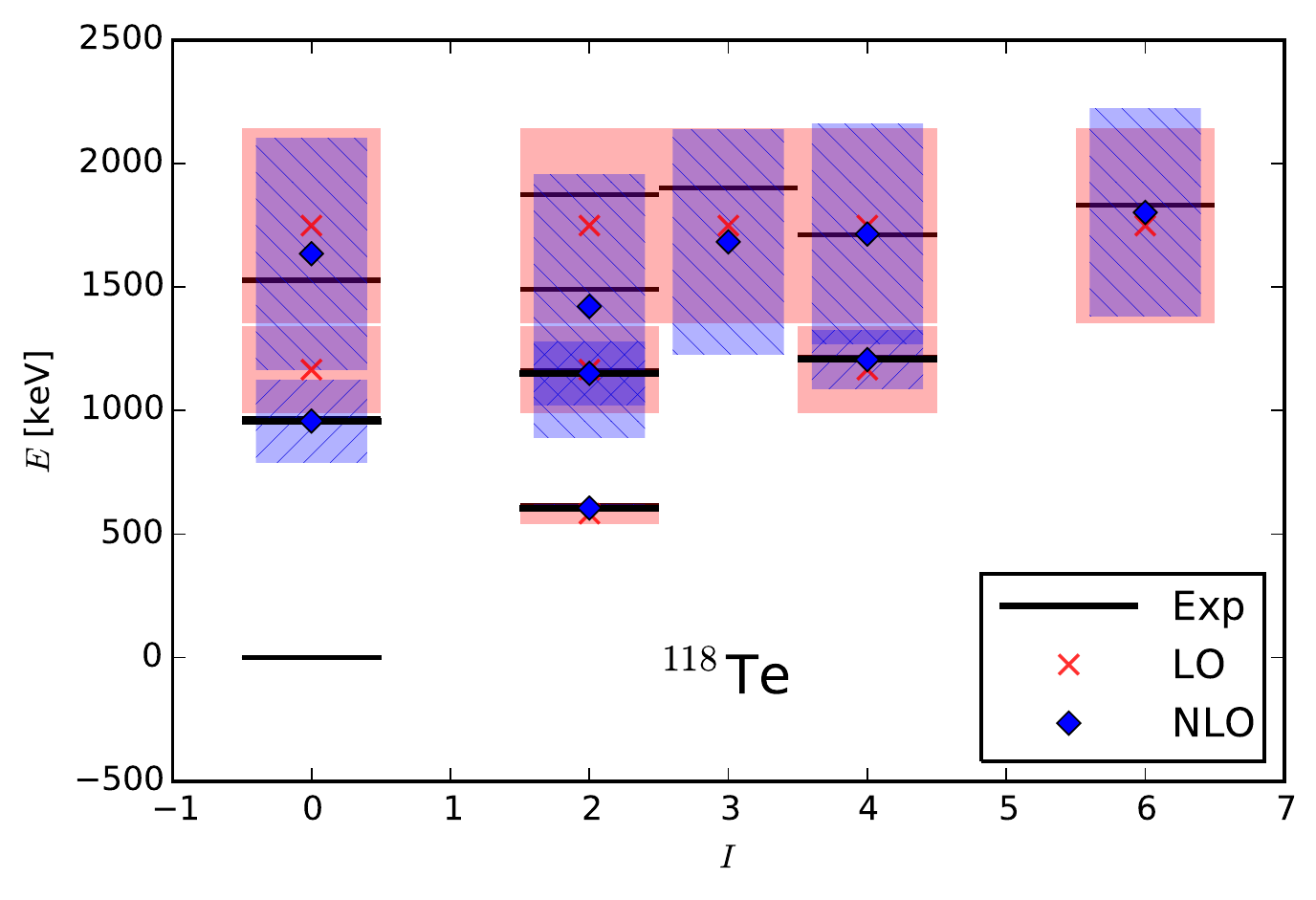}
\includegraphics[width=0.45\textwidth]{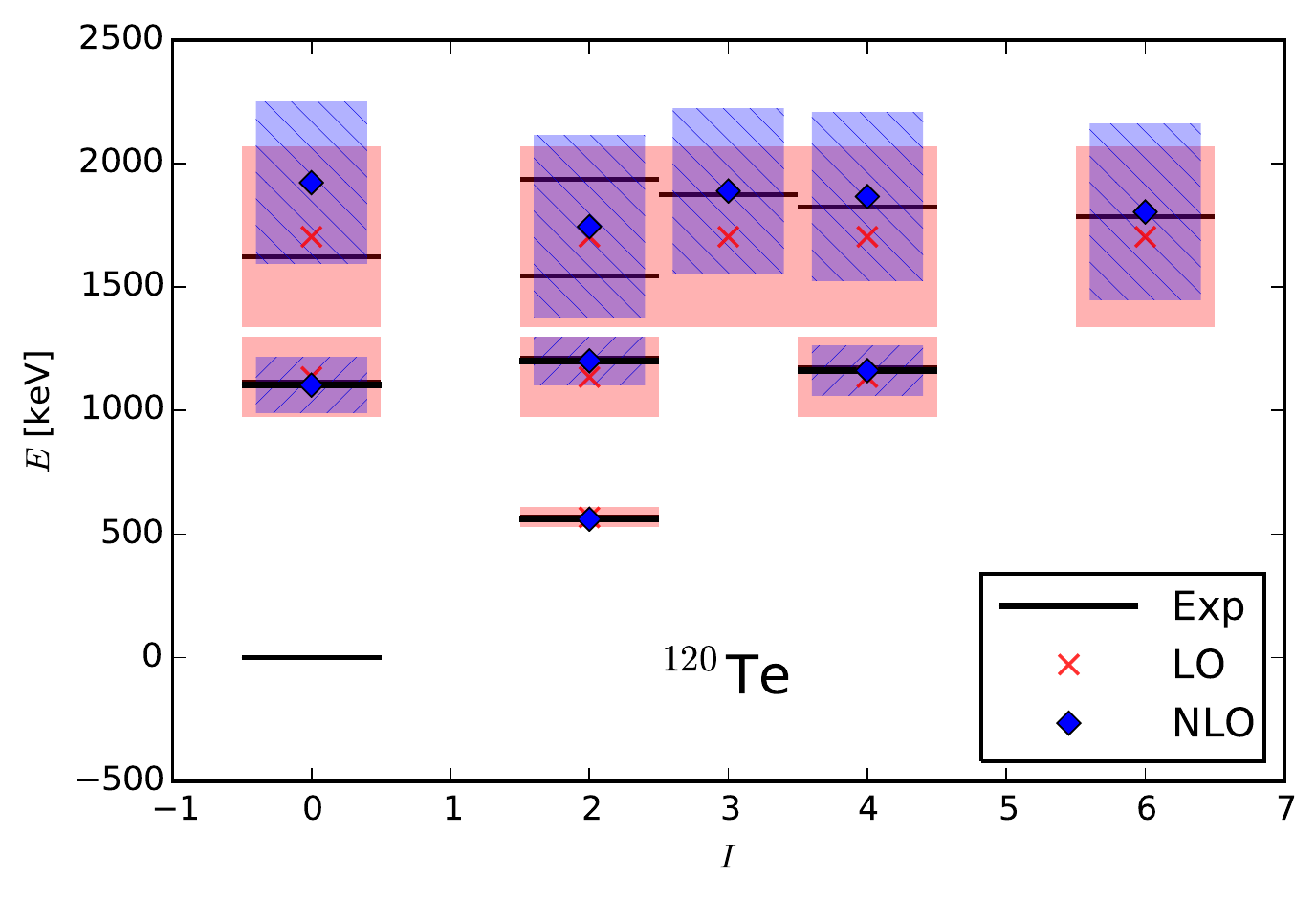}
\includegraphics[width=0.45\textwidth]{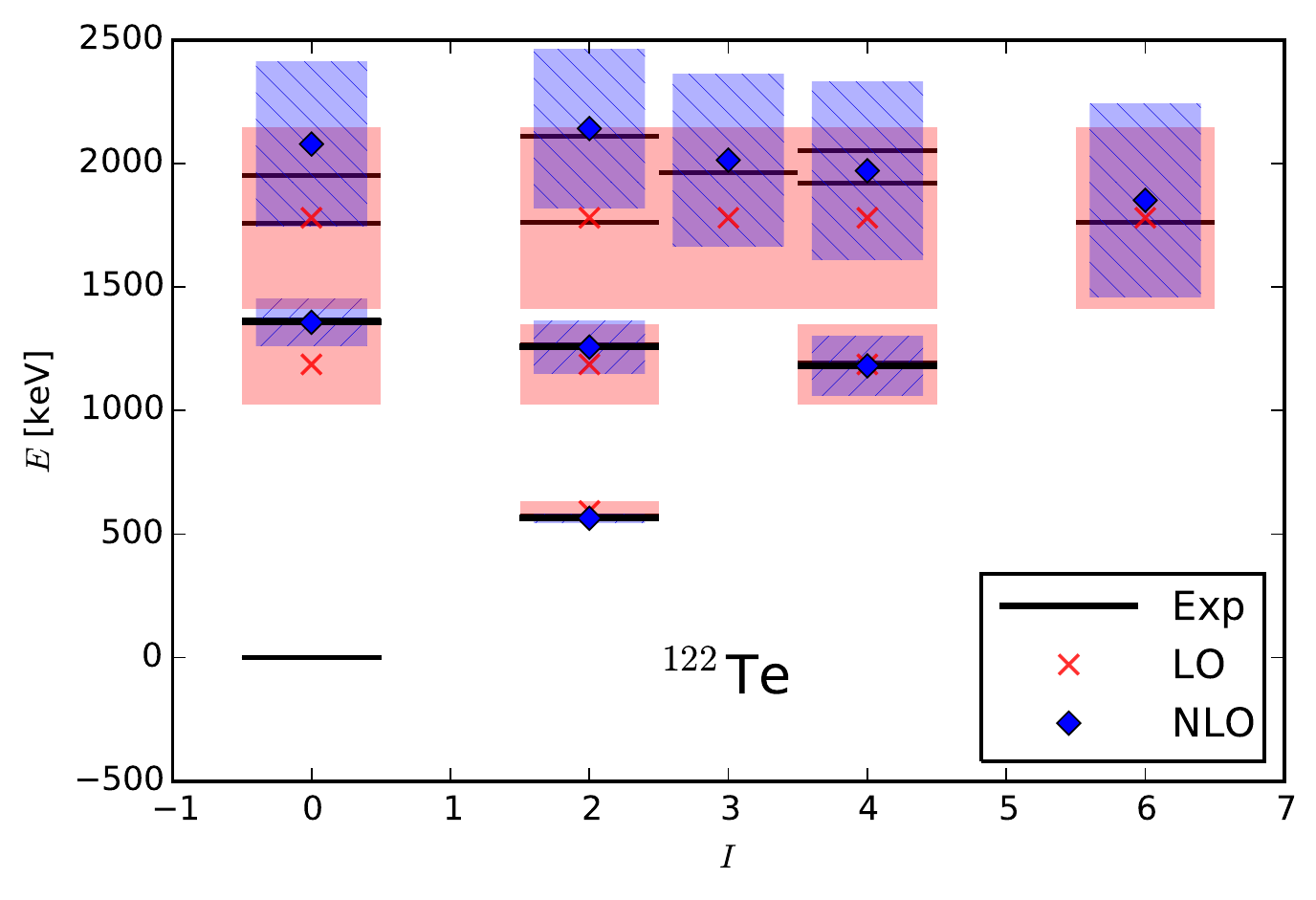}
\caption{Partial energy spectrum of $^{118}$Te, $^{120}$Te (middle) and
  $^{122}$Te (bottom) up to the three-phonon level. Experimental
  data~\cite{kitao2002, tamura2007} is compared to LO and NLO
  calculations of the effective theory. States up to the two-phonon
  level are shown as thick black lines. Theoretical uncertainties are
  shown as 68\% DOB intervals.}
\label{te}
\end{figure}

The cadmium isotopes have once been considered textbook candidates of
low-energy vibrational behavior based only on their energy
spectra~\cite{bohr1975, kern1995, rowe2010}, despite exhibiting
intruder states due to protons promoted across the $Z = 50$ shell gap
around the two-phonon level~\cite{heyde1982, meyer1977}. Other studies
on cadmium isotopes~\cite{corminboeuf2000-1, kadi2003, garrett2007,
  garrett2008, garrett2010, garrett2012} in which mixing between
vibrational and non-vibrational states is taken into account, cannot
accurately describe the electromagnetic properties of multi-phonon
candidates. They set the breakdown of vibrational behavior at the 3-
or two-phonon level depending on the isotope, and suggest a
quasi-rotational character for the low-lying excitations, based on the
large quadrupole moments of some yrast states~\cite{stone2005,
  garrett2010}. For the three isotopes studied in this work, $A=110,
112,114$, experimental data was taken from Refs.~\cite{gurdal2012,
  defrenne1996, blachot2012} respectively. The lowest $0^{+}$ and
$2^{+}$ states above the one-phonon level were employed as the
two-phonon states for the $\chi^{2}$ fits. The states identified as
members of two-phonon triplet in this work might be in disagreement
with previous studies~\cite{garrett2007, garrett2008, garrett2010},
where, for example, the $0_{2}^{+}$ in $^{112}$Cd have been identified
as an intruder state~\cite{heyde1982, wood1992}. Here, the
identification is made based on the assumption that non-vibrational
modes require more energy to be excited. As we discuss in
Sect.~\ref{Transitions}, $B(E2)$ values for decays from the identified
states seems to be in better agreement with the EFT expectations than
those from other states,

Figure~\ref{te} shows the comparison between experimental data taken
from Refs.~\cite{kitao2002, tamura2007} for $^{120}$ Te and $^{122}$Te
respectively, and LO and NLO results from EFT.

The tellurium isotopic chain provide us with some candidates to
low-energy vibrational behavior. The isotopes with $A=118, 120, 122$
all exhibit very similar spectra with states that can be identified
with those of a quadrupole vibrator up to the three-phonon level. From
these isotopes, the best candidate is $^{120}$Te with a
non-vibrational state slightly above the three-phonon
quintuplet. $^{122}$Te exhibits a non-vibrational state already at the
three-phonon level. The breakdown of the collective behavior is a
consequence of competing single-particle motion, know to exist in
tellurium isotopes~\citep{paul1995, paul1996, vanhoy2003, hicks2005,
  hicks2008, nag2012}, and signaled in $^{122}$Te by the unusual
energy ratios $E(4_{1}^{+})/E(2_{1}^{+})<2$ and
$E(6_{1}^{+})/E(4_{1}^{+})<1.5$~\cite{cizewski1989}. The alignment of
both valence nucleons and protons promoted across the $Z=50$ shell gap
breaks the spherical symmetry and give rise to non-collective deformed
states. These states compete energetically with the collective
states. In particular, the $6_{1}^{+}$ state have been interpreted
both as a vibrational state or in terms of valence protons
configurations coupled to a tin core.

Let us summarize our uncertainties as 68\% DOB intervals $\pm\delta$
for the hard-wall (hw) prior and the Gaussian (G) prior.  The
uncertainty is $\omega\delta$ for the energy levels.  At LO, the pdfs
in Eqs.~(\ref{hwLO}) and~(\ref{GLO}) agree with each other and yield
values of $\delta=0.07$ and $\delta=0.29$ for the one- and two-phonon
levels, respectively. At NLO Table~\ref{deltas} summarizes the values
of $\delta$ for states up to the two-phonon level. The columns labeled
by hw and G show the values of $\delta$ obtained from the pdfs in
Eqs.~(\ref{hwNLO}) and~(\ref{GNLO}), respectively. With the exception
of a few relatively large uncertainties, both priors yield very
similar results. For large uncertainties $\delta$, one samples the
tails of the respective priors, and these are notably different (and
not well constrained by data, cf. Fig.~\ref{cumulative}).

\begin{table}
\caption{Values of the uncertainties at NLO, with $\pm\delta$ giving
  the size of 68\% DOB intervals in states up to the two-phonon level.
  The hard-wall (hw) and Gaussian (G) priors we employed the
  distribution functions~(\ref{hwNLO}) and~(\ref{GNLO}),
  respectively.}  \centering
\begin{tabular}{c|cc|cc|cc|cc}
\hline\hline
 & \multicolumn{2}{c}{$2_{1}^{+}$} & 
	\multicolumn{2}{c}{$0_{2}^{+}$} & 
	\multicolumn{2}{c}{$2_{2}^{+}$} & 
	\multicolumn{2}{c}{$4_{1}^{+}$} \\
Nucleus & hw & G & hw & G & hw & G & hw & G \\
\hline
$^{62}$Ni & 0.02 & 0.02 & 0.29 & 0.22 & 0.21 & 0.20 & 0.20 & 0.20 \\
$^{98}$Ru & 0.02 & 0.02 & 0.18 & 0.19 & 0.18 & 0.18 & 0.18 & 0.18 \\
$^{100}$Ru & 0.04 & 0.03 & 0.18 & 0.18 & 0.30 & 0.22 & 0.21 & 0.20 \\
$^{106}$Pd & 0.03 & 0.02 & 0.18 & 0.18 & 0.18 & 0.18 & 0.21 & 0.20 \\
$^{108}$Pd & 0.02 & 0.02 & 0.18 & 0.19 & 0.18 & 0.18 & 0.18 & 0.19 \\
$^{110}$Cd & 0.02 & 0.02 & 0.18 & 0.18 & 0.18 & 0.18 & 0.19 & 0.19 \\
$^{112}$Cd & 0.02 & 0.02 & 0.18 & 0.18 & 0.18 & 0.18 & 0.18 & 0.19 \\
$^{114}$Cd & 0.02 & 0.02 & 0.18 & 0.18 & 0.18 & 0.18 & 0.18 & 0.19 \\
$^{118}$Te & 0.02 & 0.02 & 0.34 & 0.23 & 0.21 & 0.20 & 0.19 & 0.19 \\
$^{120}$Te & 0.02 & 0.02 & 0.19 & 0.19 & 0.18 & 0.18 & 0.18 & 0.19 \\
$^{122}$Te & 0.03 & 0.03 & 0.18 & 0.18 & 0.18 & 0.19 & 0.21 & 0.20 \\
\hline\hline
\end{tabular}
\label{deltas}
\end{table}

\section{Electromagnetic moments -- comparison with data}
\label{Transitions}

In this Section, we compare our results for transition quadrupole
moments, diagonal quadrupole matrix elements, and magnetic moments
with data. Theoretical uncertainties are quantified for all quadrupole
observables we consider.  As we will see, the EFT correctly captures
and consistently describes the main experimental features of
vibrational nuclei.

To determine the LEC $Q_0$ we perform $\chi^{2}$ fits to data at LO
with
\begin{equation}
\chi^2_{\rm LO}=\sum_t \frac{\left[B(E2)_{\rm exp}^{(t)}-B(E2)_{\rm
    LO}^{(t)}\right]^2} {\sigma_{\rm exp}^2 +\sigma_{\rm LO}^2} .
\end{equation}
Here $t$ labels the transitions from the one-phonon state to the
ground state and from the two-phonon states to the one-phonon state,
i.e.  $2_1^+\to 0_1^+$, $0_2^+\to 2_1^+$, $2_2^+\to 2_1^+$, and
$4_1^+\to 2_1^+$. In these fits we estimate the theoretical
uncertainty for decays from the $N$-phonon state as
\begin{equation}
\sigma_{\rm LO} =  B(E2)_{\rm LO}^{(t)}\varepsilon \ .
\end{equation}

Experimental data was mostly taken from the Nuclear Data Sheets for
the studied nuclei. For $^{62}$Ni, this data was complemented with
that from Ref.~\cite{chakraborty2011}, while for $^{98}$Ru we took the
data from Ref.~\cite{radeck2012}, which establish a ratio
$B(E2,4_{1}^{+}\rightarrow 2_{1}^{+})/B(E2,2_{1}^{+}\rightarrow
0_{1}^{+})=1.86(16)$ in agreement with the expectations for vibrators
instead of taking data for which this ratio has anomalous
values~\cite{kharraja1999, cakirli2004, williams2006}. The lack of
experimental data for $^{118}$Te makes it impossible to perform
$\chi^{2}$ fit. For $^{120}$Te, we fixed $Q_{0}^{2}$ to the only
experimental value, and make predictions for decays from the
two-phonon states.

Table~\ref{transitions} compares experimental and theoretical $B(E2)$
values (in Weisskopf units) for each nucleus considered in this work.
The theoretical uncertainty is shown as 68\% DOB intervals from the
pdf~(\ref{GLO}) with $s=1$.  Within the often considerable theoretical
uncertainties, the EFT consistently describes the available
experimental data. These results, taken together with the results for
energy level in Table~\ref{deltas}, show that vibrational nuclei can
be described as such within an EFT with a breakdown scale around the
three-phonon level. They are examples for anharmonic quadrupole
oscillators.

\begin{table}
\caption{$B(E2)$ values (in Weisskopf units) for decays from states
  below the three-phonon level in the ensemble of all studied
  nuclei. is in agreement with LO calculations below the breakdown
  level. The theoretical uncertainty is given by the 68\% DOB interval
  for the normalized residual for $B(E2)$ values. States with
  transition strengths within theoretical uncertainty are
  characterized as collective vibrations.}  \centering
  \begin{tabular}{c|ll|llll}
    \hline\hline
    Nucleus & $2_{1}^{+}\rightarrow 0_{1}^{+}$ & EFT & $0_{2}^{+}\rightarrow 2_{1}^{+}$ & $2_{2}^{+}\rightarrow 2_{1}^{+}$ & $4_{1}^{+}\rightarrow 2_{1}^{+}$ & EFT \\
    \hline 
    $^{62}$Ni & 12.1(4) & 11(4) & 42(23) & 14.9(42) & 21(6) & 21(7) \\
    $^{98}$Ru & 31(1) & 28(9) & & 47(5) & 57.6(40) & 56(19) \\
    $^{100}$Ru & 35.6(4) & 24(8) & 35(5) & 30.9(4) & 51(4) & 47(16) \\
    $^{106}$Pd & 44.3(15) & 30(10) & 35(8) & 44(4) & 76(11) & 61(20) \\
    $^{108}$Pd & 49.5(13) & 37(12) & 52(5) & 71(5) & 73(8) & 74(25) \\
    $^{110}$Cd & 27.0(8) & 21(7) & & 30(5) & 42(9) & 42(14) \\
    $^{112}$Cd & 30.2(3) & 23(8) & 51(14) & 15(3) & 61(6) & 46(15) \\
    $^{114}$Cd & 31.1(19) & 22(7) & 27.4(17) & 22(6) & 62(4) & 43(15) \\
    $^{120}$Te & 31 (6) & 31(10) & & & & 62(21) \\
    $^{122}$Te & 36.9(3) & 41(14) & & 100(30) &  & 81(27) \\
    \hline\hline
  \end{tabular}
  \label{transitions}
\end{table}

\begin{figure}[h!]
\centering
\includegraphics[width=0.45\textwidth]{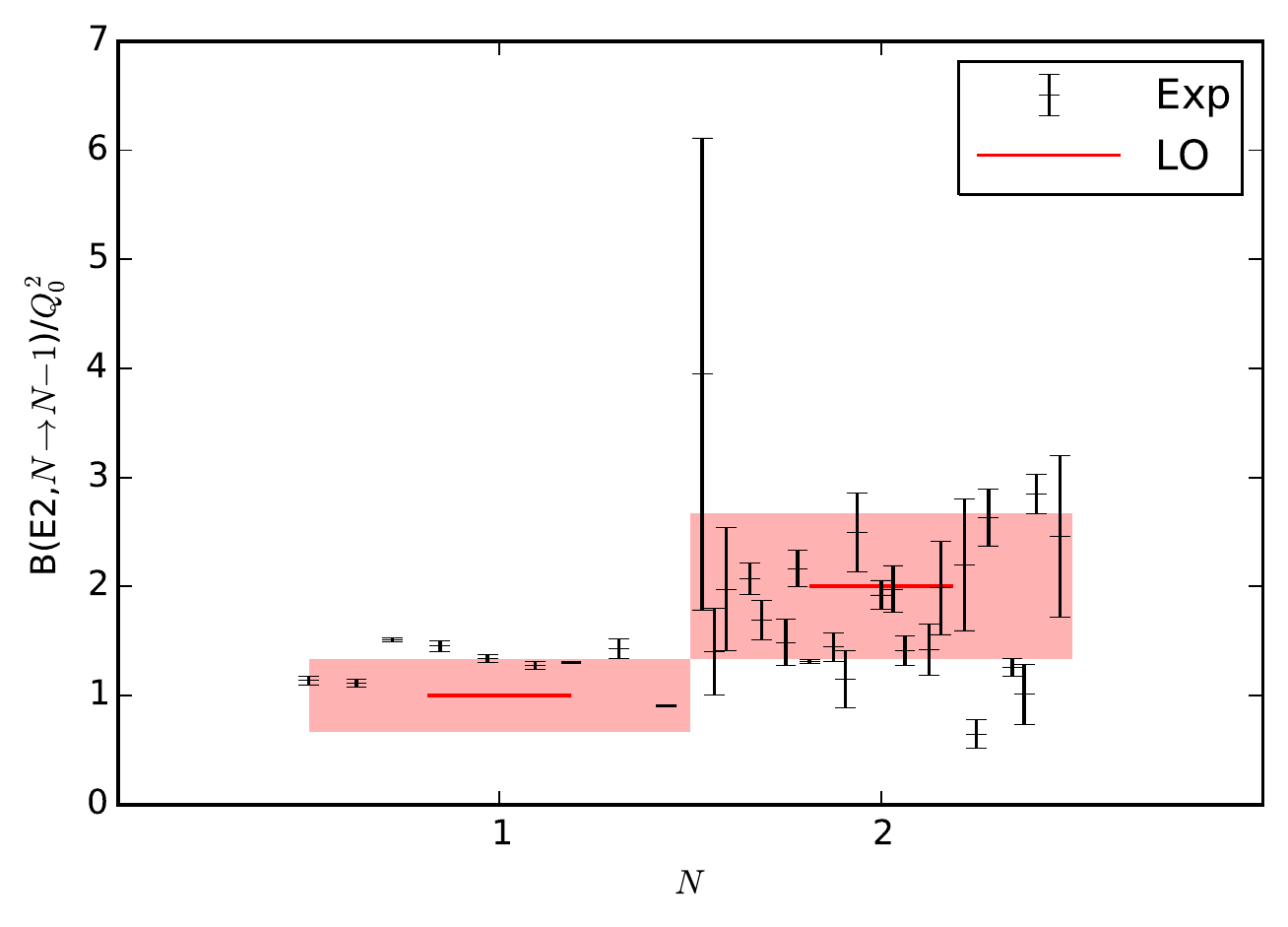}
\caption{Comparison between the normalized $B(E2)$ values for decays
  from the one- and two-phonon states in the ensemble of the nuclei
  studied in this work. Experimental $B(E2)$ values are shown as black
  lines with error bars. Quantified theoretical uncertainties are
  shown as shaded areas.}
\label{ensembleBE2}
\end{figure}

How reasonable and consistent are the 68\% DOB intervals for the
$B(E2)$ transitions?  To address this question, we turn again to the
ensemble of vibrational nuclei considered in this work.  Excluding the
isotopes $^{118,120}$Te, the EFT prediction $B(E2)/Q_0^2=N$ for decays
from the $N$-phonon state can be compared to the data from all nuclei
in the ensemble.  This comparison is shown in Fig.~\ref{ensembleBE2},
where the experimental data and the LO calculations are shown as black
errorbars and red lines with shaded uncertainty bands, respectively.
About 81\% of the data is within the 68\% DOB intervals. This is a
consistent agreement for an ensemble of 32 data points.

The eigenstates of a harmonic quadrupole oscillator have vanishing
diagonal quadrupole matrix elements. Compared to this ideal case,
diagonal quadrupole matrix elements for isotopes of Cd and Pd exhibit
sizes that are only somewhat smaller than transition quadrupole
moments.  From the EFT's perspective, sizeable diagonal quadrupole
matrix elements are expected. Comparing the expansion of the
spectrum~(\ref{ENLO}) with that of the quadrupole
operator~(\ref{quadmomexpand_d}) shows that anharmonic corrections
have relative size $\varepsilon$ for energies and relative size
$\varepsilon^{1/2}$ for the quadrupole operator.

Let us consider diagonal quadrupole matrix elements~(\ref{staticQ}).
We employ experimental data for the diagonal quadrupole matrix
elements of the $2_{1}^{+}$, $2_{2}^{+}$ and $4_{1}^{+}$ in $^{106}$Pd
and $^{108}$Pd from \citeauthor{svensson1995} and determine the LEC
$Q_1$ by a $\chi^{2}$ fit to these data. In these fits, the
theoretical uncertainty was estimated as $Q_{0}\varepsilon^{3/2}$ as
discussed in Subsection~\ref{EMcoup}.

The fits yield $Q_{1}=-0.14 \ e\rm b$ for both palladium isotopes.
(We recall that for a nucleus with $A$ nucleons $ 1~\rm{W.~U.} = 5.94
\times 10^{-6} A^{4/3}~{\rm e}^{2}\rm{b}^{2}$.)  Comparing the size of
$Q_{1}$ against $Q_{0}$ yields $Q_{1}/Q_{0}=0.47$ and
$Q_{1}/Q_{0}=0.41$ in $^{106}$Pd and $^{108}$Pd, respectively. These
ratios are consistent with the EFT estimate
$Q_{1}/Q_{0}\sim\varepsilon^{1/2}=\sqrt{1/3}\approx 0.58$. In other words,
sizeable diagonal quadrupole matrix elements are not a surprise for
these anharmonic vibrators but rather expected and due to the marginal
separation of scale, i.e. the breakdown of the EFT around the
three-phonon level.

\begin{figure}[h!]
\centering
\includegraphics[width=0.49\textwidth]{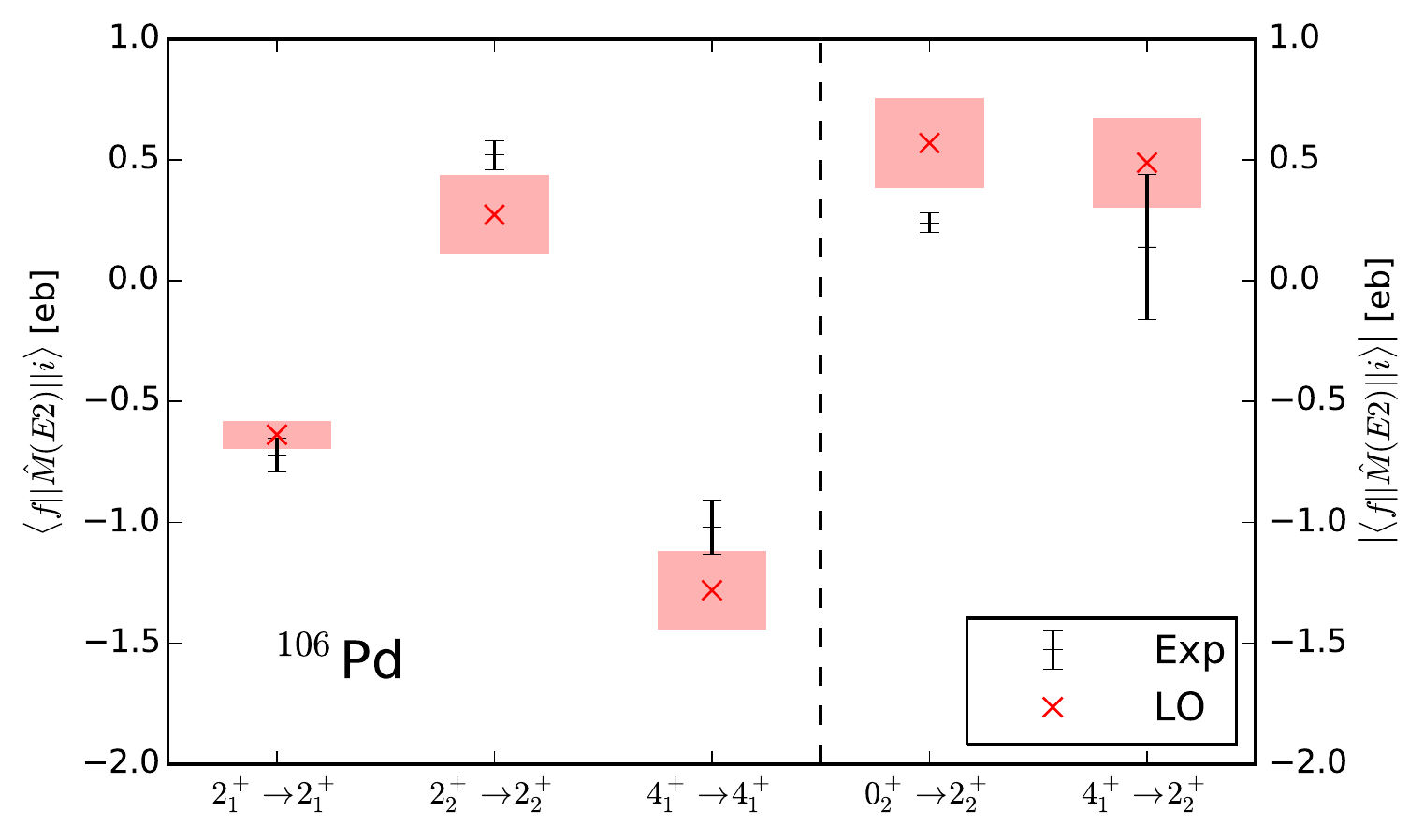}
\includegraphics[width=0.49\textwidth]{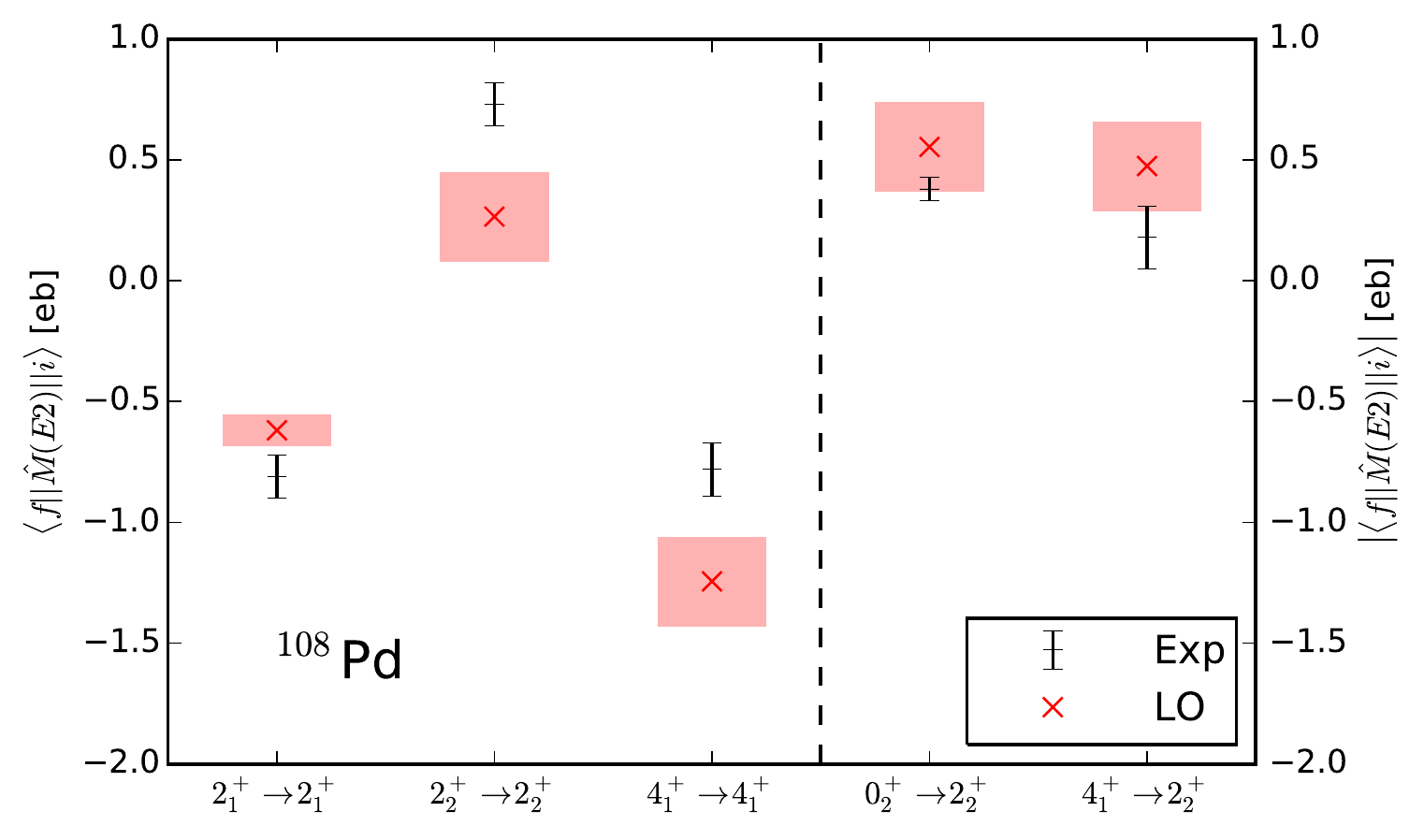}
\caption{Comparison between data and EFT results for the reduced
  quadrupole matrix elements for in $^{106}$Pd (top) and $^{108}$Pd
  (bottom). Experimental data are shown as black lines, while EFT
  results from LO calculations are shown as red diamonds with
  uncertainties as shaded 68\% DOB intervals. The left part shows
  diagonal quadrupole matrix elements employed in the fit of the LEC
  constant $Q_1$. The right part shows predictions for the absolute
  values of the reduced quadrupole matrix elements governing $E2$
  transitions between two-phonon states.}
\label{NLOquadmoments}
\end{figure}

The left part of both panels in Figure~\ref{NLOquadmoments} compares
EFT results to data~\cite{svensson1995} for the diagonal quadrupole
matrix elements of the $2_{1}^{+}$, $2_{2}^{+}$ and $4_{1}^{+}$ states
in $^{106}$Pd (top) and $^{108}$Pd (bottom). Theoretical uncertainty
are shown as 68\% DOB bands. They are based on the Gaussian
prior~(\ref{prior3}) and $M=1$ in Eq.~(\ref{res13}). Within the
theoretical uncertainties, the EFT is consistent with the data.

We turn to transition quadrupole moments~(\ref{Qtrans2phonon}) between
two-phonon states because these are also determined by the LEC $Q_1$
and are thus predictions of the EFT. The right part of
Fig.~\ref{NLOquadmoments} shows the magnitude of the transition matrix
elements and compares them to data~\cite{svensson1995}. We note that
the EFT yields different signs of these (non-observable) matrix
elements and that only the magnitude of these matrix elements
is an observable quantity, see the definition of the observable 
$B(E2)$ transition strength in Eq.~(\ref{BE2}). 

\begin{figure}[h!]
\centering
\includegraphics[width=0.49\textwidth]{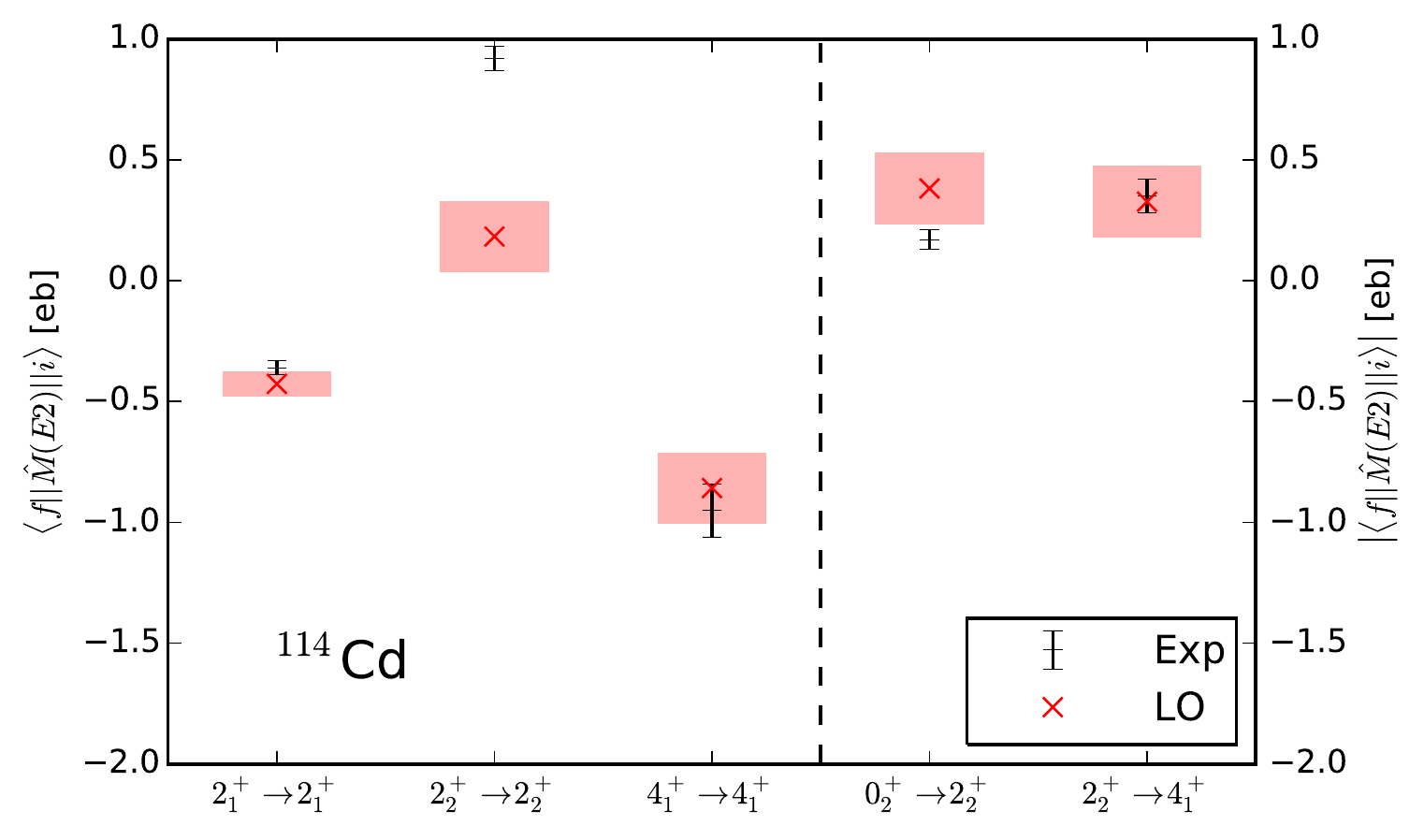}
\caption{Comparison between data and EFT results for the reduced
  quadrupole matrix elements for in $^{114}$Cd. Experimental data are
  shown as black lines, while EFT results from LO calculations are
  shown as red diamonds with uncertainties as shaded 68\% DOB
  intervals. The left part shows diagonal quadrupole matrix elements
  employed in the fit of the LEC constant $Q_1$. The right part shows
  predictions for the absolute values of the reduced quadrupole matrix
  elements governing $E2$ transitions between two-phonon states.}
\label{fig_Cd114_quad}
\end{figure}

Theoretical results for quadrupole matrix elements in $^{114}$Cd are
shown in Fig.~\ref{fig_Cd114_quad} and compared to
data~\cite{fahlander1988}. The uncertainties are quantified as for the
palladium isotopes. With the exception of the diagonal matrix element
of the $2_2^+$ state, the EFT yields a consistent description of the
data, and has predictive power for the off-diagonal matrix
elements. Here, $Q_0 = 0.27$~eb, and $Q_1 = -0.09$~eb.

Thus, the EFT consistently describes matrix elements of
electromagnetic operators. In the present approach, the
anharmonicities are due to the operators themselves, with states being
the eigenstates of the harmonic quadrupole oscillator.  We note that
Figs.~\ref{NLOquadmoments} and \ref{fig_Cd114_quad} exhibit very
similar patterns for the different nuclei.  As a last consistency
check, we turn to magnetic moments.

The EFT needs one magnetic moment to determine an LEC, i.e. the
constant $g$ in Eq.~(\ref{M1}).  While magnetic moments are typically
known for the lowest $2^+$ state in many even-even
nuclei~\cite{stone2005}, the EFT can only be tested if more magnetic
moments are known below the three-phonon level. The states $2_1^+$,
$2_2^+$, and $4_1^+$ have non-zero spins and thus exhibit magnetic
moments. As discussed below Eq.~(\ref{M1}), the EFT predicts at LO
that both $2^+$ states have equal magnetic moments, i.e.
$\mu(2_1^+)=\mu(2_2^+)\equiv\mu(2^+)$, and that the $4^+$ state has a
magnetic moment $\mu(4_1^+)=\sqrt{6}\mu(2^+)\approx
2.44\mu(2+)$. Weighted averages of the experimental
data~\cite{stone2005} [in units of nuclear magnetons (nm)] for
$^{106}$Pd shows that $\mu(2_1^+)\approx 0.79\pm 0.02$~nm,
$\mu(2_2^+)=0.71\pm 0.10$~nm, and $\mu(4_1^+)=1.8\pm 0.4$~nm.  This is
consistent with EFT expectations. It would certainly be interesting to
test these EFT predictions in other vibrational nuclei.

Overall, the EFTs results and predictions for electromagnetic
properties of states and transitions below the three-phonon level are
consistent with data. This would make it interesting to measure such
complete data sets for other vibrational nuclei as well.

\section{Summary}
\label{Summary}
We developed an EFT for collective nuclear vibrations based on
quadrupole degrees of freedom, rotational invariance, and a breakdown
scale at around the three-phonon level.  For spectra, the EFT is
driven to next-to-leading order, while the computation of other matrix
elments is restricted to leading order. The terms in appearing in the
Hamiltonian and quadrupole operator differ from those employed in
several models. 

The EFT approach also allows us to quantify theoretical
uncertainties. To this purpose, we make testable assumptions about
priors regarding the distribution of low-energy constants and employ
recently developed tools from Bayesian statistics. We give analytical
results for the important case of log-normal priors. The priors
employed in the uncertainty quantification of energies are consistent
for the ensemble of nuclei we considered.

The EFT is minimally coupled to electromagnetic gauge fields in a
model-independent way, with non-minimal couplings accounting for
subleading corrections. For states below the three-phonon level we
describe LO $B(E2)$ transition strengths with quantified uncertainties
and present several results for diagonal and off-diagonal matrix
elements of the quadrupole operator.  Comparing the EFT results to an
extensive data set shows that spectra and transition strengths are
consistently described within the theoretical and experimental
uncertainties for $^{62}$Ni, $^{98,100}$Ru, $^{106,108}$Pd,
$^{110,112,114}$Cd, and $^{118,120,122}$Te.  In particular, relatively
large diagonal matrix elements in $^{106,108}$Pd and $^{114}$Cd are
consistent with the expectations of the EFT. The consistent
description of spectra, $E2$ transitions and matrix elements, and
magnetic moments within the EFT for nuclear vibration suggests that
the nuclei studied in this work can be viewed as anharmonic quadrupole
vibrators. This work also suggests that it would be interesting to
measure a combination of matrix elements for electric and magnetic
observables in nuclei such as $^{120}$Te and $^{122}$Te.

It would be interesting to extend the EFT of nuclear vibrations also
to odd-mass neighbors of the even-even nuclei considered in this
work. Combining, for instance, halo EFT with this work, one might
explore to what extent such nuclei can be understood by coupling the
odd nucleon to the quadrupole degrees of freedom of vibrational
even-even nuclei.

\begin{acknowledgments}
This material is based upon work supported by the U.S.  Department of
Energy, Office of Science, Office of Nuclear Physics under Award
Number DEFG02-96ER40963 (University of Tennessee), and under Contract
No.  DE-AC05-00OR22725 (Oak Ridge National Laboratory).
\end{acknowledgments}

\begin{appendices}

\section{Analytical results for log-normal priors}
\label{AppUQ}
In this Appendix we present some details for the derivation of
analytical results for the combination of log-normal
priors~(\ref{prior1}) and hard-wall priors~(\ref{prior2}).

The denominator of Eq.~(\ref{master}) is
\bea
\lefteqn{\int\limits_0^\infty {\rm d}c \,\,{\rm pr}(c) \prod\limits_{m=0}^k {\rm pr}^{(\rm hw)}(c_m|c) =}\nonumber\\
& {2^{-(k+1)}\over\sqrt{2\pi}\sigma}\int\limits_a^\infty {\rm d} c\,\, c^{-(k+2)} e^{-{1\over 2\sigma^2}(\log c)^2} \ .
\eea
Here,
\be
a\equiv \max(|c_0|,\ldots,|c_k|)
\ee
is a function of the expansion coefficients. Substitutions $z=\log c$
and $x=z-\log(a)$ yield 
\be
\frac{e^{-(k+1)\log a} e^{-{1\over 2\sigma^2}(\log a)^2}}
{2^{(k+1)}\sqrt{2\pi}\sigma}
\int\limits_0^\infty {\rm d} x\,\, e^{-{x^2\over 2\sigma^2}-x\left(k+1+{\log a\over\sigma^2}\right)} \ . 
\ee
This integral is known~\cite{gradshteyn}, and we find
\be 
2^{-(k+2)} e^{{\sigma^2\over 2}(k+1)^2} \left[
  1-\Phi\left({\sigma\over\sqrt{2}}\left(k+1 +{\log a\over
    \sigma^2}\right)\right)\right] 
\ee 
as the final result for the denominator of Eq.~(\ref{master}).  Here,
$\Phi(x)\equiv(2/\sqrt{\pi})\int_0^x {\rm d}t \exp{(-t^2)}$ denotes
the error function.  The numerator of the expression~(\ref{master})
can be evaluated in similar fashion. Employing the shorthand
\be
b\equiv \max\left(a,{|\Delta|\over \varepsilon^{k+1}}\right)
\ee
we find for the numerator of Eq.~(\ref{master})
\be
{e^{{\sigma^2\over 2}(k+2)^2} \over 2^{k+3}\varepsilon^{k+1}} 
\left[ 1-\Phi\left({\sigma\over\sqrt{2}}\left(k+2 +{\log b\over \sigma^2}\right) \right)\right] \ .
\ee
Thus, for $M=1$
\be
\label{result1app}
p_1^{(\rm hw)}(\Delta|c_0,\ldots,c_k)=
{e^{{2k+3\over 2}\sigma^2}\over 2 \varepsilon^{k+1}}
\frac
{1-\Phi\left({\sigma\over\sqrt{2}}\left(k+2 +{\log b\over \sigma^2}\right) \right)}
{1-\Phi\left({\sigma\over\sqrt{2}}\left(k+1 +{\log a\over \sigma^2}\right)\right)} \ , \nonumber 
\ee
and the dependence on the expansion coefficients is entirely contained
in the functions $a$ and $b$.

Let us continue and compute $p_2^{(\rm hw)}(\Delta|c)$. The integral~(\ref{p}) is
again known for $M=2$~\cite{gradshteyn}, and the final result is
\bea
\label{pM2}
p_2^{(\rm hw)}(\Delta|c) =\left\{\begin{array}{ll}
{1\over 2 \varepsilon^{k+1}c} \ ,  & |\Delta| \le (1-\varepsilon)\varepsilon^{k+1}c\\
0 \ , & |\Delta|> (1+\varepsilon)\varepsilon^{k+1}c\\
{(1+\varepsilon)\varepsilon^{k+1}c-|\Delta|\over 4 \varepsilon^{2k+3}c^2}\ ,  &\mbox{else}
\end{array}\right.\nonumber
\eea
As we need to integrate over $c$ for the computation of
$p_2^{(\rm hw)}(\Delta|C_0,\ldots,c_k)$, we rewrite this function as
\bea
p_2^{(\rm hw)}(\Delta|c) =\left\{\begin{array}{ll}
0 & \quad\mbox{for}\quad c\le{|\Delta|\over (1+\varepsilon)\varepsilon^{k+1}}\\
{1\over 2\varepsilon^{k+1}c} &\quad\mbox{for}\quad c> {|\Delta|\over (1-\varepsilon)\varepsilon^{k+1}}\\
{(1+\varepsilon)\varepsilon^{k+1}c-|\Delta|\over 4 \varepsilon^{2k+3}c^2} &\quad\mbox{else} 
\end{array}\right.\nonumber
\eea
The remaining integrations are similar to the ones solved above, and one finds
\begin{widetext}
\bea
\label{result2}
p_2^{(\rm hw)}(\Delta|c_0,\ldots,c_k)&=&
\frac{ \left(2 \varepsilon^{k+1}\right)^{-1} e^{{2k+3\over 2}\sigma^2} }
{1-\Phi\left({\sigma\over\sqrt{2}}\left(k+1 +{\log a\over \sigma^2}\right)\right)}
\Bigg\{
1-\Phi\left({\sigma\over\sqrt{2}}\left(k+2 +{\log d\over \sigma^2}\right) \right) \nonumber\\
&&+
{1+\varepsilon\over 2 \varepsilon}\Theta(g-f)\left[
\Phi\left({\sigma\over\sqrt{2}}\left(k+2 +{\log g\over \sigma^2}\right) \right)
-\Phi\left({\sigma\over\sqrt{2}}\left(k+2 +{\log f\over \sigma^2}\right) \right)\right]
\nonumber\\
&&-{|\Delta|\over 2 \varepsilon^{k+2}}\Theta(g-f)e^{{2k+5\over 2}\sigma^2}\left[
\Phi\left({\sigma\over\sqrt{2}}\left(k+3 +{\log g\over \sigma^2}\right) \right)
-\Phi\left({\sigma\over\sqrt{2}}\left(k+3 +{\log f\over \sigma^2}\right) \right)\right]
\Bigg\}
\eea
\end{widetext}

Here, $\Theta$ denotes the unit step function, and the expressions
\bea 
d&\equiv& \max\left(a,{|\Delta|\over (1-\varepsilon)\varepsilon^{k+1}}\right) \ ,\\ 
f&\equiv& \max\left(a,{|\Delta|\over (1+\varepsilon)\varepsilon^{k+1}}\right) \ ,\\ 
g&\equiv& {|\Delta|\over (1-\varepsilon)\varepsilon^{k+1}} 
\eea 
encode much of the functional dependence.

For $M>2$, the evaluation of $p_M^{(\rm hw)}(\Delta|c)$
[Eq.~(\ref{p})] becomes increasingly tedious. Fortunately, $p_2^{(\rm
  hw)}(\Delta|c)$ is a good approximation even for $M>2$. The quality
of this approximation can be verified by inserting the
expression~(\ref{p}) into Eq.~(\ref{master}) and performing the
integrations numerically.  We note that the accuracy of the $M=2$
result is not surprising. As the expansion coefficients $c_n$ are
natural in size, increasingly higher orders contribute little to the
residual~(\ref{residual}). This makes Eq.~(\ref{result2}) the main
result of this Appendix.

\end{appendices}

\end{document}